\documentclass[lettersize,journal]{IEEEtran}
\usepackage{amsmath,amsfonts}
\usepackage{array}
\usepackage[caption=false,font=normalsize,labelfont=sf,textfont=sf]{subfig}
\usepackage{textcomp}
\usepackage{stfloats}
\usepackage{url}
\usepackage{verbatim}
\usepackage{graphicx}
\usepackage{cite}
\usepackage{amssymb}
\usepackage{caption}
\usepackage[ruled,norelsize]{algorithm2e}

\makeatletter
\newcommand{\removelatexerror}{\let\@latex@error\@gobble}
\makeatother

\usepackage{booktabs}
\usepackage{multirow}
\usepackage{color}
\usepackage{tikz}
\newcommand*\circled[1]{%
  \tikz[baseline=(char.base)]{
    \node[shape=circle,fill,inner sep=0.7pt] (char)
      {\fontsize{8pt}{8pt}\selectfont\textcolor{white}{#1}};}}
      
\usepackage[hidelinks]{hyperref}
\newcommand{\ignore}[1]{}

\begin{document}

\title{BackWeak: Backdooring Knowledge Distillation \\Simply with Weak Triggers and Fine-tuning}

\author{Shanmin~Wang\IEEEauthorrefmark{1},
        and~Dongdong~Zhao\thanks{\IEEEauthorrefmark{2} Corresponding author: Dongdong Zhao (zdd@whut.edu.cn).}\IEEEauthorrefmark{1}\IEEEauthorrefmark{2}
        
\IEEEauthorblockA{\IEEEauthorrefmark{1}School of Computer Science and Artificial Intelligence, Wuhan University of Technology}}

\markboth{Journal of \LaTeX\ Class Files,~Vol.~14, No.~8, August~2021}%
{Shell \MakeLowercase{\textit{et al.}}: A Sample Article Using IEEEtran.cls for IEEE Journals}

\maketitle

\begin{abstract}
Knowledge Distillation (KD) is essential for compressing large models, yet relying on pre-trained ``teacher'' models downloaded from third-party repositories introduces serious security risks—most notably backdoor attacks. Existing KD backdoor methods are typically complex and computationally intensive: they employ surrogate student models and simulated distillation to guarantee transferability, and construct triggers similar to universal adversarial perturbations (UAPs), which being not stealthy in magnitude, inherently exhibit strong adversarial behavior. This work questions whether such complexity is necessary and constructs stealthy ``weak'' triggers—imperceptible perturbations that have negligible adversarial effect. We propose BackWeak, a simple, surrogate-free attack paradigm. BackWeak shows that a powerful backdoor can be implanted by simply fine-tuning a benign teacher with a weak trigger using a very small learning rate. We demonstrate that this delicate fine-tuning is sufficient to embed a backdoor that reliably transfers to diverse student architectures during a victim's standard distillation process, yielding high attack success rates. Extensive empirical evaluations on multiple datasets, model architectures, and KD methods show that BackWeak is efficient, simpler, and often more stealthy than previous elaborate approaches. This work calls on researchers studying KD backdoor attacks to pay particular attention to the trigger's potential adversarial characteristics. 
\end{abstract}


\section{Introduction}

\IEEEPARstart{D}{eep} neural networks (DNNs) have achieved state-of-the-art performance across many tasks, but powerful architectures and weights are often computationally and memory intensive~\cite{han2015deep}. To deploy such models under resource constraints, model compression techniques such as knowledge distillation (KD)~\cite{gou2021knowledge,yang2023categories,moslemi2024survey} are widely adopted. KD transfers knowledge from a large, high-performance ``teacher'' model to a compact ``student'', preserving much of the teacher's accuracy while reducing computational and memory demands. In practice, practitioners usually obtain pretrained teacher models and weights from third-party repositories (e.g., HuggingFace, Zenodo) and apply KD to adapt those models to their needs. However, relying on externally sourced models greatly expands the attack surface: pretrained artifacts that pass repository-level static security scans~\cite{huggingface_hub_security} may nevertheless cause unwanted or malicious effects owing to insufficient auditing of the runtime behaviors encoded in model weights, and the black-box nature of DNNs makes it feasible for an adversary to embed such bad behaviors such as backdoor attacks~\cite{gu2017badnets,nguyen2021wanet,wang2022bppattack}. 

Backdoor attacks aim to implant a clandestine mapping between an input-space trigger and a target output, remaining dormant under clean inputs but being activated upon trigger presence~\cite{li2022backdoor,gao2020backdoor}. The feasibility and impact of backdoors have been widely demonstrated across domains~\cite{gu2017badnets,cheng2025backdoor,yan2024backdoor}. In the context of KD, recent work~\cite{ge2021anti,chen2025taught} has explored paradigms that simulate the distillation process with surrogate student models to craft transferable backdoors. Crucially, a common element of these techniques is an explicit trigger-optimization stage, these triggers are commonly shaped by cross-entropy objectives that compel the model to predict a target class for any input stamped with the trigger, yielding triggers that resemble \emph{universal adversarial perturbations} (UAPs)~\cite{moosavi2017universal}, we therefore refer to them as \emph{UAP-like} triggers. Although effective, these approaches heavily rely on the intrinsic misleading capability of the UAP-like triggers. To achieve strong attack effectiveness, these triggers often need to be less constrained, which makes them visually noticeable, while also possessing the capability to perform adversarial attacks themselves.

Although~\cite{chen2025taught} defines the perturbation magnitude constraint using $\epsilon_0$, it does not provide an evaluation for it. In~\cite{ge2021anti}, the trigger pixel values are constrained by $\frac{1}{c}$, and different values of $c$ are evaluated; however, even the most constrained trigger they evaluated (with $c=10$) remains visually noticeable and exhibits strong adversarial effect. For Universal Adversarial Perturbations (UAPs), it is evident that stronger perturbations generally have a greater ability to mislead models into misclassification. 
These observations suggest that the effectiveness of prior KD backdoor attacks may stem less from a genuinely implanted backdoor and more from the strong adversarial nature of the UAP-like triggers themselves. If the trigger's adversarial effect is already doing most of the work, then a computationally expensive simulated distillation stage analogous to that in \cite{ge2021anti} becomes largely unnecessary, and the attack may effectively degenerate into generating a UAP-like trigger and training the model to associate it with the target label. However, experimental findings from~\cite{liu2023ink} show that traditional approaches which train the backdoor task jointly with the benign task~\cite{nguyen2021wanet,zeng2023narcissus} tend to exhibit weak transferability through distillation. We suppose that this is because the backdoor task becomes effectively decoupled from the benign task under large learning rates, causing the model's outputs on clean samples to carry almost no backdoor-related signal, leaving little exploitable bias for the student to inherit during distillation.

This motivates a further question: \emph{instead of embedding the backdoor jointly with the benign task during training, could we first train a benign model and then fine-tune it with a carefully controlled strategy so that the backdoor becomes coupled with the benign task---gently influencing the model's outputs on clean samples and thereby enhancing its transferability?} Going one step further, \emph{could such a backdoor be constructed using an even more covert and less adversarial trigger?}

Surprisingly, our answer to the above questions is affirmative. We propose \textbf{BackWeak}, a surrogate-free and lightweight attack flow that \circled{1} synthesizes a \emph{weak trigger} with a benign model and then \circled{2} implants a backdoor into the teacher model by fine-tuning with the trigger under a small learning rate. The resulting teacher retains benign accuracy while embedding a backdoor that is transferable to the distilled students. Here, a \emph{weak trigger} denotes a perturbation whose adversarial effect on a benign model is negligible and is usually visually stealthy.

We empirically validate BackWeak across multiple architectures, datasets and KD methods. Our evaluation shows that \textbf{(i)} weak triggers crafted as above can produce high attack success rates after distillation despite being nearly imperceptible; \textbf{(ii)} the workflow is simpler than prior surrogate-based attacks while still achieving strong performance; and \textbf{(iii)} BackWeak exhibits better stealthiness under standard perceptual and normed metrics. Importantly, we analyze the inherent adversarial properties of triggers employed by representative prior works (e.g., ADBA~\cite{ge2021anti} and SCAR~\cite{chen2025taught}).

Our contributions are as follows:
\begin{itemize}
  \item For the first time, we identify and demonstrate that a backdoor learned via low-learning-rate, teacher-only fine-tuning with \emph{weak} triggers reliably transfers through KD to students, eliminating the need for surrogate students and distillation simulation. This substantially lowers attack cost and increases practical risk.
  \item We carefully analyze the adversarial properties of triggers used in prior work and show that many of these triggers function more as strong adversarial perturbations than as genuine backdoor triggers, thereby clarifying that their surrogate-student and simulated-distillation procedures are far less necessary than previously assumed.
  \item We conduct extensive experiments over multiple datasets, architectures and KD methods, validating BackWeak's simplicity, effectiveness, and stealthiness. 
\end{itemize}

We hope this work draws the community's attention to backdoor threats in knowledge-distillation settings and, in particular, motivates thorough investigation of the intrinsic adversarial properties of triggers when studying such attacks. Our code is available at \href{https://github.com/SomeBottle/8ackW3ak}{\emph{GitHub}}.

\section{Background and Related Work}

\subsection{Knowledge Distillation}
\label{subsec:knowledge_distillation}

\subsubsection{Overview}

Knowledge distillation (KD) transfers inter-class similarities and calibrated uncertainty of a high-capacity teacher to a compact student by encouraging the student to match the teacher's predictive distribution~\cite{gou2021knowledge}. This approach compresses models and reduces inference cost with acceptable loss in accuracy, making it suitable for resource-constrained scenarios. In the canonical, logit-based method (vanilla KD)~\cite{hinton2015distilling}, the student \(f_{\theta_S}\) is trained on a convex combination of hard-label cross-entropy and a KL divergence term between temperature-softened distributions:  
\begin{equation}
\label{eq:distill}
\begin{aligned}
\min_{\theta_S}\ &\mathbb{E}_{(\boldsymbol{x},y)\sim\mathcal{P}}\!\left[
(1-\alpha)\cdot\mathcal{L}_{\mathrm{CE}}\!\big(\mathbf{z}_S(\boldsymbol{x}),\ y\big) \right.
\\
&\left. +\ \alpha\cdot\tau^{2}\cdot
\mathcal{L}_\mathrm{KL}\!\big(\mathbf{p}_T^\tau(\boldsymbol{x})\,\|\,\mathbf{p}_S^\tau(\boldsymbol{x})\big)
\right],
\end{aligned}
\end{equation}

\noindent where \(\alpha\!\in\![0,1]\) balances the two losses, \(\tau\ge 1\) controls the softness via \(p^\tau_i\!=\exp(z_i / \tau) / \sum_j \exp(z_j / \tau)\), \(\mathbf{z}\) denotes the logit vector (e.g., \(\mathbf{z}_S(\boldsymbol{x})\) or \(\mathbf{z}_T(\boldsymbol{x})\)), \(i\) indexes classes, \(\mathcal{P}\) is a data distribution, and the factor \(\tau^2\) keeps gradient magnitudes comparable as \(\tau\) varies. Larger \(\tau\) produces a flatter teacher distribution that exposes class-similarity structure beyond one-hot targets.

\subsubsection{Taxonomy and Recent Work}

Since its introduction by Hinton~et~al.~\cite{hinton2015distilling}, knowledge distillation (KD) has evolved into a diverse field, with methods now broadly categorized into three main paradigms: \emph{response-based}, \emph{feature-based}, and \emph{relation-based} approaches~\cite{gou2021knowledge,yang2023categories,moslemi2024survey}.   

\textit{Response-based} methods~\cite{hinton2015distilling,song2023exploring,kim2023rckd,song2025flexible} are generally vanilla KD and its variants, which align the output distribution of teacher and student models. \textit{Feature-based} methods~\cite{romero2014fitnets,chen2022knowledge,ji2021show,yang2024vitkd} transfer knowledge by compelling the student to mimic the teacher's intermediate feature activations, while \textit{relation-based} methods~\cite{park2019relational,huang2022knowledge,xin2024new} focus on preserving structural relationships between data samples or feature maps. 

Most KD methods, including the one in Equation~\ref{eq:distill}, formulate the student's objective $\mathcal{L}_S$ as a linear combination of the supervised cross-entropy (CE) loss $\mathcal{L}_{\mathrm{CE}}$ and the distillation loss $\mathcal{L}_{\mathrm{KD}}$, i.e., $\mathcal{L}_{S} = \alpha \mathcal{L}_{\mathrm{KD}} + \beta \mathcal{L}_{\mathrm{CE}}$. The response-based, feature-based, and relation-based paradigms differ primarily in how $\mathcal{L}_{\mathrm{KD}}$ is defined.

While feature- and relation-level objectives can yield task- or architecture-specific gains, vanilla KD remains a strong, stable baseline across diverse settings (especially on large-scale data~\cite{hao2023revisit}) and is simple to implement~\cite{sanh2019distilbert,zhang2025shiftkd}. For these reasons, vanilla KD is widely adopted.

\subsection{Universal Adversarial Perturbations}

\subsubsection{Overview}

Moosavi-Dezfooli~et~al.\cite{moosavi2017universal}\ first demonstrated the existence of universal adversarial perturbations (UAPs)---single, input-agnostic vectors that cause trained classifiers to misclassify a large fraction of inputs---thereby inspiring extensive research on adversarial robustness~\cite{cohen2019certified,xiong2024all}.

Formally, let \(f:\mathbb{R}^d\to\mathbb{R}^C\) denote a classifier with prediction function \(k(\boldsymbol{x})=\arg\max f(\boldsymbol{x})\). Given a data distribution \(\mathcal{P}\) and a tolerance parameter \(\delta\in(0,1)\), a perturbation vector \(\boldsymbol{v}\in\mathbb{R}^d\) is said to be universal if
\begin{equation}
\begin{aligned}
\Pr_{\boldsymbol{x}\sim\mathcal{P}_{\mathcal{X}}}\!\left[k(\boldsymbol{x}+\boldsymbol{v})\neq k(\boldsymbol{x})\right] &\ge 1-\delta, \\
\text{s.t.}\ \|\boldsymbol{v}\|_p \le \varepsilon,\  \boldsymbol{x}+\boldsymbol{v} &\in [0,1]^d,
\end{aligned}
\label{eq:uap}
\end{equation}

\noindent where $\varepsilon$ denotes the perturbation budget under an $\ell_p$-norm constraint, ensuring that $\boldsymbol{v}$ remains imperceptible to human observers. In empirical settings, the probability above is estimated via the fooling rate over a finite dataset.

\subsubsection{Characteristics}  

The defining characteristic of UAPs is their \emph{universality}~\cite{moosavi2017universal}: a single perturbation is effective across a distribution of inputs rather than a single instance. Beyond this, UAPs exhibit notable \emph{transferability} across architectures\cite{chaubey2020universal,hashemi2020transferable,zhang2025improving}, whereby a perturbation generated for one model often retains effectiveness against models with distinct architectures or training strategies. \ignore{These phenomena suggest that decision boundaries of high-dimensional classifiers admit shared geometric structures and low-dimensional adversarial subspaces.}  

In this paper, we focus on the characteristics of UAPs and carefully evaluate the adversarial properties of the triggers constructed in prior work as well as those produced by our own method.

\begin{figure*}[!t]
\centering
\captionsetup{font=normalsize}
\includegraphics[width=0.8\textwidth]{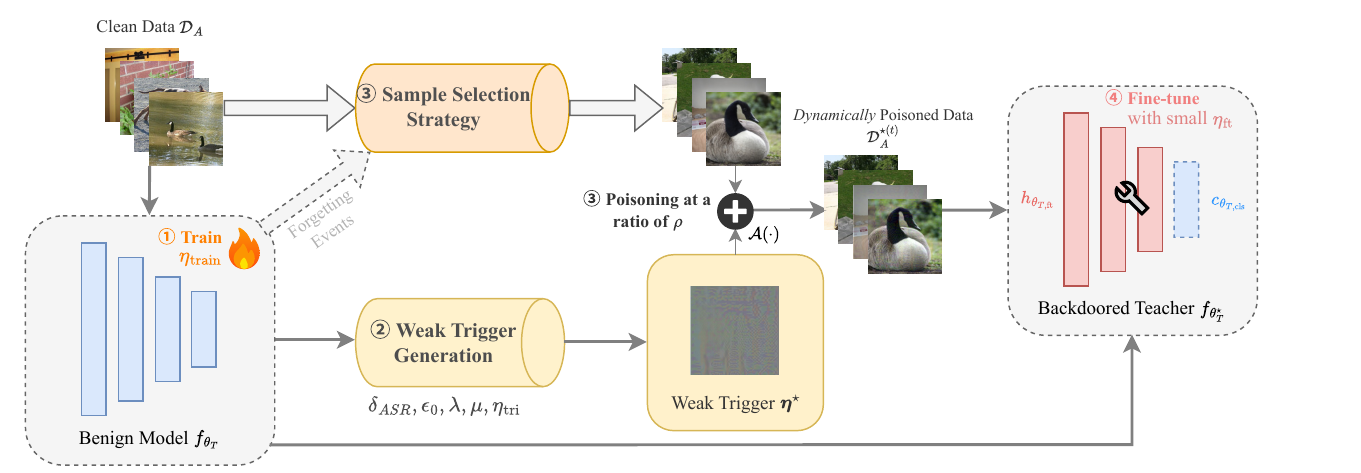}
\caption{Overview of the proposed \texttt{BackWeak} workflow. }
\label{fig:workflow_overview} 
\end{figure*}

\subsection{Backdoor Attacks}

\subsubsection{Overview}

Backdoor (a.k.a.\ Trojan) attacks implant an \emph{attacker-chosen} behavior into a model that remains dormant on clean inputs but is activated whenever a \emph{trigger} appears. Since BadNets~\cite{gu2017badnets}, most studies have focused on \emph{targeted} attacks in classification tasks~\cite{li2022backdoor,gao2020backdoor}. Formally, let \(\mathcal{K}_\theta:\mathcal{X}\!\to\!\mathcal{Y}\) be a classifier and \(\mathcal{P}\) a distribution over \(\mathcal{X}\!\times\!\mathcal{Y}\). The adversary chooses a trigger pattern \(\boldsymbol{\eta}\) and target label \(y_t\!\in\!\mathcal{Y}\), injects malicious behavior via \emph{data poisoning}, \emph{training-time manipulation}, or \emph{model-level modification}, yielding parameters \(\theta^\star\) that satisfy:
\begin{equation}
\label{eq:backdoor}
\begin{aligned}
&\Pr_{(\boldsymbol{x},y)\sim\mathcal{P}}\!\big[\mathcal{K}_{\theta^\star}(\boldsymbol{x})=y\big]\ \ge 1-\varepsilon,\\
&\Pr_{\boldsymbol{x}\sim\mathcal{P}_{\mathcal{X}}}\!\big[\mathcal{K}_{\theta^\star}(\boldsymbol{x}\oplus\boldsymbol{\eta})=y_t\big]\ \ge 1-\delta,
\end{aligned}
\end{equation}
\noindent where \(\oplus\) denotes the trigger-applying operation, and the small values \(\varepsilon\) and \(\delta\) quantify the classifier's desired performance on inputs without and with the trigger, respectively.   

\subsubsection{Backdoor Triggers vs.\ UAPs}
\label{subsubsec:backdoor_triggers_vs_uaps}
A backdoor trigger should only take effect when the model was \emph{maliciously manipulated} to associate a specific pattern with the attacker predefined behavior~\cite{gu2017badnets}. In contrast, a universal adversarial perturbation (UAP) is a test-time, input-agnostic perturbation exploiting intrinsic vulnerabilities \textbf{without} tampering with the model. Therefore, there should be  
\begin{equation}
\label{eq:nobackdoor}
\begin{aligned}
\Pr_{\boldsymbol{x}\sim\mathcal{P}_{\mathcal{X}}}\!\big[\mathcal{K}_{\theta}(\boldsymbol{x}\oplus\boldsymbol{\eta})=y_t\big]\ \le\ \delta',
\end{aligned}
\end{equation}

\noindent where $\delta'$ is small, indicating that a benign model \textbf{should not} strongly respond to the trigger with the behavior.   

\subsubsection{UAPs as Backdoor Triggers}  

Despite the distinctions, UAPs have been adopted as backdoor triggers by many researchers. For instance, UAPs have been employed to enhance backdoor attacks in model compression~\cite{phan2022invisible} and to implement attacks in model extraction~\cite{wang2025honeypotnet}. Their application also extends to generative models, as seen in I2I~\cite{jiang2024backdoor} and UIBDiffusion~\cite{han2025uibdiffusion}. However, several other studies have generated UAP-like patterns as triggers without explicitly identifying them as such (e.g., PatchBackdoor~\cite{yuan2023patchbackdoor}, BadMerging~\cite{zhang2024badmerging}, and ADBA~\cite{ge2021anti}), so it remains debatable whether they can be considered backdoor attacks.

\subsubsection{Backdooring Knowledge Distillation}

Ge~et~al.~\cite{ge2021anti} pioneered backdoor attacks in the knowledge distillation scenario with ADBA, establishing a paradigm based on surrogate student models. Subsequently, Wang~et~al.~\cite{wang2025honeypotnet} proposed the similar HoneypotNet for model extraction scenarios, and Cheng~et~al.~\cite{cheng2024transferring} adapted this attack to LLMs with ATBA. In concurrent work, Chen~et~al.'s SCAR~\cite{chen2025taught} also built upon this paradigm, enhancing the backdoor's stealthiness within the teacher model. Other paradigms include attacks based on intrinsic dataset properties~\cite{liu2023ink}, those targeting feature distillation~\cite{chen2024like}, and those directly poison the distillation set~\cite{wu2025backdoor}.

Attacks utilizing surrogate models exhibit considerable generality. However, focusing on image classification tasks, existing methods in this paradigm \circled{1} simulate the distillation process, incurring conspicuous computational overhead, and \circled{2} generate UAP-like patterns as triggers, yet may lack ablation studies on the trigger's effect on benign models. Therefore, we propose a direct, surrogate-free attack workflow against KD and investigate whether a UAP-based trigger implants a genuine backdoor or functions merely as an adversarial perturbation.

\section{Threat Model}

We formalize the threat model by defining the attacker's objectives and the capabilities they possess, grounded in a practical scenario of decentralized model distribution.

\subsection{Attacker's Goal}
\label{subsec:goal}

The attacker's goal is to construct a backdoored ``teacher'' model and disseminate it via public, third-party model repositories (e.g., Hugging Face, GitHub, Zenodo). This scenario exploits a practical vulnerability: such platforms often lack the resources for rigorous runtime security auditing of the vast number of uploaded model files, while users may simultaneously lack the capability to verify their integrity. 

An honest downstream practitioner (the victim), seeking to adapt a large pre-trained model for resource-constrained devices (such as mobile phones or IoT devices), may download this compromised model and use it for knowledge distillation (KD). The attacker's primary goal is to ensure the backdoor, implanted during the teacher's training, is effectively and automatically transferred to the ``student'' model during the victim's distillation process. This transferred backdoor must remain stealthy---causing no significant degradation on benign inputs---while reliably forcing any input containing the attacker-specified trigger to be classified as the attacker's chosen target class.

\subsection{Attacker's Capability}
\label{subsec:capability}

Similar to prior work~\cite{ge2021anti,wang2025honeypotnet,chen2025taught,chen2024like}, we assume the attacker operates in a white-box setting relative to the teacher model, possessing full control over its architecture, training data, optimization process, and all associated hyperparameters. 

However, the attacker's capabilities are severely limited regarding the victim's distillation phase. The attacker publishes the model without any knowledge of who will download it or their specific implementation. Therefore, the attacker has \textbf{zero knowledge} of the student model's architecture, the specific KD strategy (e.g., loss functions, temperature) employed, or the hyperparameters used during the victim's training. Critically, we posit a realistic scenario where the attacker does not have access to the victim's dataset used for distillation. This assumption distinguishes our model from prior work (e.g., ADBA~\cite{ge2021anti}, ATBA~\cite{chen2024like}), which often presumed a shared dataset. We only assume that the attacker's and victim's datasets, while disjoint, are drawn from a similar underlying data distribution (we also discuss the scenario under a distribution shift in Sec.\ref{subsec:robustness_to_distribution_shift}). In summary, the attacker is only capable of training the malicious teacher model and releasing its weights.

\begin{table}[htbp]
\centering
\caption{Important notations.}
\label{tab:notations}
\begin{tabular}{ll}
\toprule
Notation & Definition \\
\midrule
\multicolumn{2}{l}{\emph{Models and Data}} \\
$f_{\theta_T}$, $f_{\theta_S}$ & Teacher model, student model. \\
$f_{\theta_T^\star}$ & Final malicious (backdoored) teacher model. \\
$h_{\theta_{T, \mathrm{ft}}}$ & Feature extractor part of the teacher model. \\
$c_{\theta_{T, \mathrm{cls}}}$ & Classification layer of the teacher model. \\
$\mathcal{X}$, $\mathcal{Y}$ & Input space and label space (C classes). \\
$\mathcal{D}_A$, $\mathcal{D}_V$ & Attacker's dataset, victim's (disjoint) dataset. \\
$\mathcal{P}$ & Underlying data distribution shared by $\mathcal{D}_A$, $\mathcal{D}_V$. \\
$y_t$ & Target class for the attack. \\
$k(\cdot)$ & Prediction function ($\arg\max$). \\
\\
\multicolumn{2}{l}{\emph{Triggers and Poisoning}} \\
$G(\cdot)$ & Trigger injection / applying function. \\ 
$\boldsymbol{\eta}$, $\boldsymbol{\eta}^\star$ & Trigger, optimized weak trigger. \\
$\mathcal{C}$ & Set of valid image pixel values (e.g., $[-1, 1]^d$). \\
$\mathcal{A}(\cdot)$ & Random transforms applied to the trigger. \\
$\Pi_{\mathcal{C}}(\cdot)$ & Projection function onto the valid set $\mathcal{C}$. \\
$\mathcal{D}_A^{\star(t)}$ & Dynamically poisoned dataset for epoch $t$. \\
\\
\multicolumn{2}{l}{\emph{Hyperparameters}} \\
$\delta_{\text{ASR}}$ & ASR budget for weak trigger generation. \\
$\rho$ & Poisoning ratio. \\
$\epsilon_0$ & $\ell_\infty$-norm constraint for the trigger. \\
$\lambda$ & Weighting hyperparameter for $\mathcal{L}_{\text{margin}}$. \\
$\mu$ & Margin hyperparameter for $\mathcal{L}_{\text{margin}}$. \\
$\eta_{\mathrm{train}}$ & Learning rate for benign model training. \\
$\eta_{\mathrm{tri}}$ & Learning rate for weak trigger generation. \\
$\eta_{\mathrm{ft}}$ & Reduced learning rate for backdoor fine-tuning. \\
\bottomrule
\end{tabular}
\end{table}

\section{Methodology}

In this section, we first formalize the problem of backdoor transferability in knowledge distillation within our defined threat model. We then present the complete workflow of our proposed \texttt{BackWeak} attack, detailing its key stages. The key notations are listed in Table~\ref{tab:notations}.

\subsection{Problem Formulation}

Let $f_{\theta_T}:\mathcal{X}\to\mathbb{R}^C$ be the teacher classification model with parameters $\theta_T$, mapping an input $\boldsymbol{x} \in \mathcal{X}$ to a logit vector $\mathbf{z}_T\in\mathbb{R}^C$, $f_{\theta_S}:\mathcal{X}\to\mathbb{R}^C$ be the student model with parameters $\theta_S$, and $k: \mathbb{R}^C\to\mathcal{Y}$ be the prediction function defined as $k(\mathbf{z}) = \arg\max \mathbf{z}$. The attacker possesses a training dataset $\mathcal{D}_A=\{(\boldsymbol{x}_i, y_i)\}_{i=1}^{N_A}$, while the victim possesses an independent, disjoint distillation dataset $\mathcal{D}_V=\{(\boldsymbol{x}_j, y_j)\}_{j=1}^{N_V}$, where $\mathcal{D}_A\cap\mathcal{D}_V=\varnothing$. Here we assume $\mathcal{D}_A$ and $\mathcal{D}_V$ are drawn from the same underlying data distribution $\mathcal{P}$ on $\mathcal{X} \times \mathcal{Y}$.

The attacker selects a target class $y_t \in \mathcal{Y} = \{1, \dots, C\}$ and defines a trigger-applying function $G: \mathcal{X} \to \mathcal{X}$, which embeds the trigger into an input $\boldsymbol{x}$. %

\subsubsection{Definition of Weak Triggers}
\label{subsubsec:weak_trigger_def}

Formally, we call a trigger \emph{weak} if, for a benign model $f_\theta: \mathcal{X}\!\to\!\mathbb{R}^C$ trained on $\mathcal{P}$, applying it to inputs from $\mathcal{P}_{\mathcal{X}}$ induces only a negligible change in the model's predictive behavior towards a specific target class $y_t$. Let $p_0 = \Pr_{\boldsymbol{x}\sim\mathcal{P}_{\mathcal{X}}}\!\big[k(f_\theta(\boldsymbol{x}))=y_t\big]$ be the baseline probability of predicting $y_t$ on clean inputs. The trigger is considered weak if:
\begin{equation}
\label{eq:weak_trigger}
\Pr_{\boldsymbol{x}\sim\mathcal{P}_{\mathcal{X}}}\!\big[k(f_\theta(G(\boldsymbol{x})))=y_t\big] \le p_0 + \epsilon_t,
\end{equation}
\noindent where $\epsilon_t$ denotes a small tolerance margin (e.g., $\epsilon_t \le 0.1$). This definition contrasts sharply with Universal Adversarial Perturbations (UAPs) (cf. Eq.~\ref{eq:uap}), which are designed to \emph{maximize} the misclassification rate on benign models. Our investigation centers on whether such weak triggers, which have minimal effect on benign models, can be weaponized for backdoor attacks.

\subsubsection{Attacker's Objective}

The attacker's goal is to produce a malicious teacher model $f_{\theta_T^\star}$ such that any student $f_{\theta_S^\star}$, trained by a victim via an unknown knowledge distillation process $\mathcal{L}_S$ on their private dataset $\mathcal{D}_V$ with:
\begin{equation}
\label{eq:victim_kd}
\theta_S^\star = \underset{\theta_S}{\arg\min} \mathbb{E}_{(\boldsymbol{x},y)\in\mathcal{D}_V} \left[ \mathcal{L}_S(f_{\theta_T^\star}, f_{\theta_S}, \boldsymbol{x}, y) \right],
\end{equation}
\noindent will satisfy the two primary backdoor conditions (cf. Eq.~\ref{eq:backdoor}):
\begin{enumerate}
\item\textbf{Utility (Benign Accuracy):} The student performs accurately on benign inputs from the data distribution.
$$ \Pr_{(\boldsymbol{x},y)\sim\mathcal{P}}\! \big[k(f_{\theta_S^\star}(\boldsymbol{x}))=y\big] \ge 1 - \varepsilon_S $$
\item\textbf{Effectiveness (Attack Success Rate):} The student reliably predicts $y_t$ for any input processed by the trigger-applying function $G$.
$$ \Pr_{\boldsymbol{x}\sim\mathcal{P}_\mathcal{X}}\! \big[k(f_{\theta_S^\star}(G(\boldsymbol{x})))=y_t\big] \ge 1 - \delta_S $$
\end{enumerate}
\noindent where $\varepsilon_S, \delta_S$ are small tolerance values.

According to the threat model (Sec.~\ref{subsec:capability}), the attacker cannot observe or optimize Eq.~\ref{eq:victim_kd} directly. Instead, the attacker needs to find optimal teacher parameters $\theta_T^\star$ and a trigger $\boldsymbol{\eta}^\star$ by optimizing a proxy objective on their own dataset $\mathcal{D}_A$. This objective must (1) ensure the teacher's utility to make it appealing for download, (2) embed the backdoor within the teacher's distillable knowledge so that it can transfer to $f_{\theta_S}$ via knowledge distillation, and crucially, (3) constrain the trigger to be a \emph{weak trigger} with negligible adversarial effect to inject a genuine backdoor.

Formally, the attacker's optimization problem is to find $\theta_T^\star$ and $\boldsymbol{\eta}^\star$ by minimizing a composite loss function, subject to the weak trigger constraint:

\begin{equation}
\label{eq:attacker_objective}
\begin{aligned}
\theta_T^\star, \boldsymbol{\eta}^\star = &\arg\min_{\theta_T, \boldsymbol{\eta}} \left((1-\rho) \cdot \mathcal{L}_{\mathrm{Utility}} + \rho \cdot \mathcal{L}_{\mathrm{Poison}} \right), \\
\text{s.t.}\quad &\Pr_{\boldsymbol{x}\sim\mathcal{P}_{\mathcal{X}}}\!\big[k(f_\theta(G(\boldsymbol{x})))=y_t\big] \le p_0 + \epsilon_t,
\end{aligned}
\end{equation}
\noindent where $\rho$ reflects the poisoning rate, and the constraint guarantees the trigger remains weak on a benign model $f_\theta$ (as defined in Eq.~\ref{eq:weak_trigger}). The two loss components are defined over the attacker's dataset $\mathcal{D}_A$:
\begin{itemize}
    \item \textbf{Utility Loss ($\mathcal{L}_{\mathrm{Utility}}$):} Preserves high accuracy on benign tasks.
    \item \textbf{Poisoning Loss ($\mathcal{L}_{\mathrm{Poison}}$):} Instills the backdoor behavior in a manner that is transferable via knowledge distillation. 
\end{itemize}

\subsection{Overview of BackWeak Workflow}
\label{sec:backweak_overview}

To investigate the transferability of backdoors based on weak triggers during knowledge distillation, we propose the \texttt{BackWeak} attack workflow.
Unlike prior work that simulates the distillation process and employs surrogate student models~\cite{ge2021anti,chen2025taught,wang2025honeypotnet}, we adopt a direct and simple approach.
As illustrated in Figure~\ref{fig:workflow_overview}, the \texttt{BackWeak} workflow comprises four primary stages:

\vspace{0.3em}

\noindent\mbox{\textbf{Stage \circled{1}: Model Training.}} The attacker first trains a benign teacher model $f_{\theta_T}$ on their private dataset $\mathcal{D}_A$ using a standard learning rate $\eta_{\mathrm{train}}$.
The optimization objective is:
\begin{equation}
    \min_{\theta_T} \mathbb{E}_{(\boldsymbol{x},y)\in\mathcal{D}_A} \left[ \mathcal{L}_{\mathrm{CE}}(f_{\theta_T}(\boldsymbol{x}), y) \right].
\end{equation}
During this benign training phase, the number of forgetting events for each training sample is recorded, following the method proposed by Toneva et~al.~\cite{toneva2018empirical}.

\vspace{0.3em}

\noindent\mbox{\textbf{Stage \circled{2}: Weak Trigger Generation.}} Using $f_{\theta_T}$, the attacker generates a weak trigger $\boldsymbol{\eta}^\star$ under specific constraints.
This trigger is designed to satisfy the weakness definition (cf. Eq.~\ref{eq:weak_trigger}), ensuring it does not induce significant misclassification towards the target class on the benign model.
The detailed generation process is described in Section~\ref{subsec:weak_trigger_generation}.

\vspace{0.3em}

\noindent\mbox{\textbf{Stage \circled{3}: Sample Selection and Poisoning.}} The attacker selects a subset of samples from $\mathcal{D}_A$ for poisoning, controlled by the poisoning ratio $\rho$.
We explore two selection strategies: \emph{random selection} and \emph{forgetting-events-based selection}.
With the selected samples, the attacker then constructs a \emph{dynamically} poisoned dataset $\mathcal{D}_A^{\star(t)}$ for each fine-tuning epoch $t$.
This process is detailed in Section~\ref{subsec:sample_selection_and_data_poisoning}.

\vspace{0.3em}

\noindent\mbox{\textbf{Stage \circled{4}: Backdoor Injection.}} Finally, the attacker fine-tunes the benign model $f_{\theta_T}$ on the dynamically poisoned dataset $\mathcal{D}_A^{\star(t)}$ using a \emph{smaller learning rate}.
This step injects the backdoor, yielding the final malicious teacher model $f_{\theta_T^\star}$.
This stage is detailed in Section~\ref{subsec:backdoor_injection_via_fine_tuning}.

Through this four-stage process, the attacker obtains a backdoored teacher model $f_{\theta_T^\star}$, which can then be disseminated through third-party platforms for downstream victims to use in knowledge distillation.

\subsection{Weak Trigger Generation}
\label{subsec:weak_trigger_generation}

The generation of the weak trigger $\boldsymbol{\eta}$ is formulated as a constrained optimization problem, adapting techniques from Universal Adversarial Perturbation (UAP)~\cite{moosavi2017universal} discovery. Unlike traditional UAPs which are optimized to maximize the fooling rate, our objective is to find an input-agnostic trigger $\boldsymbol{\eta}^\star$ that adheres to our weakness definition (Eq.~\ref{eq:weak_trigger}), which must be optimized to create a latent association with the target class $y_t$ while \emph{simultaneously} constraining its empirical Attack Success Rate (ASR) on the benign model $f_{\theta_T}$ to be below a predefined budget $\delta_{\text{ASR}}$. This optimization is performed over the attacker's dataset $\mathcal{D}_A$ with the model parameters $\theta_T$ held constant.

Here we define the \textbf{triggered input} as $\boldsymbol{x}'_t = G(\boldsymbol{x}) = \Pi_{\mathcal{C}}\big(\boldsymbol{x} + \boldsymbol{\eta}\big)$, where $\mathcal{C}$ denotes the set of valid image pixel values, and $\Pi_{\mathcal{C}}(\cdot)$ projects its input onto $\mathcal{C}$ to ensure that $\boldsymbol{x}'_t$ is a valid image. The optimization objective consists of two components: a ``push'' loss $\mathcal{L}_{\text{push}}$ and a ``margin'' loss $\mathcal{L}_{\text{margin}}$.

\subsubsection{Push Loss (\texorpdfstring{$\mathcal{L}_{\text{push}}$}{L\_push})}

This loss guides the optimization of $\boldsymbol{\eta}$ to create an association with the target class $y_t$. It is defined as the negative difference between the log-probabilities of the target class $y_t$ for the triggered input $\boldsymbol{x}'_t$ and the clean input $\boldsymbol{x}$:
\begin{equation}
\label{eq:push_loss}
\mathcal{L}_{\text{push}} = \mathbb{E}_{(\boldsymbol{x},y)\in\mathcal{D}_A} \left[ -(\log p(y_t | \boldsymbol{x}'_t) - \log p(y_t | \boldsymbol{x})) \right],
\end{equation}
\noindent where $p(y_t | \cdot) = \text{Softmax}(f_{\theta_T}(\cdot))_{y_t}$. Minimizing $\mathcal{L}_{\text{push}}$ w.r.t. $\boldsymbol{\eta}$ iteratively modifies the trigger to ``push'' the logit of $y_t$ higher for triggered samples, creating the desired latent association.

\subsubsection{Margin Loss (\texorpdfstring{$\mathcal{L}_{\text{margin}}$}{L\_margin})}

This loss guides the optimization of $\boldsymbol{\eta}$ to \emph{discourage} it from causing arbitrary misclassifications. The objective is to penalize the trigger $\boldsymbol{\eta}$ if it causes a misclassification where the logit of the highest incorrect class exceeds the logit of the true class by more than a specified margin $\mu$. This margin $\mu$ acts as a tolerance, allowing for small logit disturbances that do not confidently lead to a misclassification. The loss is computed only over triggered samples that are already misclassified by the model: 
\begin{equation}
\label{eq:margin_loss}
\mathcal{L}_{\text{margin}}\!=\!\mathbb{E}_{\substack{(\boldsymbol{x},y)\in\mathcal{D}_A \\ k(\boldsymbol{z}'_t) \neq y}} \left[\max\!\left(0,\max_{j \neq y}(\mathbf{z}'_t)_j\!-\!(\mathbf{z}'_t)_y\!-\!\mu \right)\!\right],
\end{equation}
\noindent where $\mathbf{z}'_t = f_{\theta_T}(\boldsymbol{x}'_t)$ is the logit vector, $\max_{j \neq y}(\mathbf{z}'_t)_j$ is the logit of the strongest incorrect class. Minimizing this loss w.r.t. $\boldsymbol{\eta}$ pushes the trigger back towards a state where it is less likely to induce such misclassifications, by attempting to keep the logit difference within the tolerance $\mu$.

\subsubsection{Constrained Optimization}
The key to generating a \emph{weak} trigger lies in conditionally optimizing for $\mathcal{L}_{\text{push}}$. The final optimization problem to find the optimal trigger $\boldsymbol{\eta}^\star$ is:
\begin{equation}
\label{eq:weak_trigger_opt}
\begin{aligned}
\boldsymbol{\eta}^\star = \underset{\boldsymbol{\eta}}{\arg\min}\ &\mathbb{I}(\text{ASR}_{\text{epoch}} < \delta_{\text{ASR}}) \cdot \mathcal{L}_{\text{push}} +\lambda \mathcal{L}_{\text{margin}}, \\
\text{s.t.}\ &\|\boldsymbol{\eta}\|_\infty \le \epsilon_0, 
\end{aligned}
\end{equation}
\noindent where $\lambda$ is a weighting hyperparameter, $\mathbb{I}(\cdot)$ is the indicator function, $\delta_{\text{ASR}}$ is the ASR budget, and $\text{ASR}_{\text{epoch}}$ is the empirical attack success rate of the trigger $\boldsymbol{\eta}$ from the previous epoch:
\begin{equation}
\text{ASR}_{\text{epoch}} = \frac{\sum_{(\boldsymbol{x},y)\in\mathcal{D}_A} \mathbb{I}(k(f_{\theta_T}(\boldsymbol{x}'_t))\!=\!y_t \land y\!\neq\!y_t)}{\sum_{(\boldsymbol{x},y)\in\mathcal{D}_A} \mathbb{I}(y\!\neq\!y_t)}
\end{equation}

Eq.~\ref{eq:weak_trigger_opt} ensures that $\boldsymbol{\eta}$ is optimized for effectiveness ($\mathcal{L}_{\text{push}}$) only when its empirical ASR remains below the budget $\delta_{\text{ASR}}$. If the ASR exceeds the budget, the optimization only minimizes $\mathcal{L}_{\text{margin}}$, effectively controlling the trigger's adversariality.  

In practice, this problem is solved iteratively using a gradient-based optimizer (e.g., SGD with a cosine annealing scheduler) with a learning rate $\eta_\mathrm{tri}$. After each update step, the $\ell_\infty$ constraint is enforced via clipping. The complete procedure is summarized in Algorithm~\ref{alg:weak_trigger_gen}.

\begin{figure}[!t]
\removelatexerror
\begin{algorithm}[H]
\label{alg:weak_trigger_gen}
\SetAlgoLined
\KwIn{Benign model $f_{\theta_T}$, dataset $\mathcal{D}_A$, target $y_t$, budget $\delta_{\text{ASR}}$, constraint $\epsilon_0$, epochs $E$, weights $\lambda, \mu$, learning rate $\eta_{\mathrm{tri}}$}
\KwOut{Weak trigger $\boldsymbol{\eta}$}
\BlankLine
Initialize $\boldsymbol{\eta} \gets \mathbf{0}$\;
$ASR_{current} \gets 0.0$\;
\For{i $ = 1$ to $E$}{
    \For{batch $(\boldsymbol{X}, \boldsymbol{y})$ in $\mathcal{D}_A$}{
        \tcp{Eq.~\ref{eq:push_loss}}
        $\mathcal{L}_{\text{push}} \gets \mathcal{L}_{\text{push}}(f_{\theta_T}, \boldsymbol{\eta}, \boldsymbol{X}, y_t)$\; 
        \tcp{Eq.~\ref{eq:margin_loss}}
        $\mathcal{L}_{\text{margin}} \gets \mathcal{L}_{\text{margin}}(f_{\theta_T}, \boldsymbol{\eta}, \boldsymbol{X}, \boldsymbol{y}, \mu)$\;
        
        $\mathcal{L}_{\text{total}} \gets \lambda \cdot \mathcal{L}_{\text{margin}}$\;
        \If{$ASR_{current} < \delta_{\text{ASR}}$}{
            $\mathcal{L}_{\text{total}} \gets \mathcal{L}_{\text{total}} + \mathcal{L}_{\text{push}}$\;
        }

        $\boldsymbol{\eta} \gets \text{GradientBasedUpdate}(\boldsymbol{\eta}, \mathcal{L}_{\text{total}}, \eta_{\mathrm{tri}})$\;
        \tcp{Project to $\ell_\infty$-ball}
        $\boldsymbol{\eta} \gets \text{clip}(\boldsymbol{\eta}, -\epsilon_0, +\epsilon_0)$
    }
    $ASR_{current} \gets \text{EvaluateASR}(\boldsymbol{\eta}, \mathcal{D}_A, y_t)$\;
}
\Return $\boldsymbol{\eta}$\;
\caption{Weak Trigger Generation Process.}
\end{algorithm}
\end{figure}

\subsection{Sample Selection and Data Poisoning}
\label{subsec:sample_selection_and_data_poisoning}

After generating the weak trigger $\boldsymbol{\eta}^\star$, the attacker selects a subset of indices $\mathcal{I}_{\text{poison}} \subset \{1, \dots, N_A\}$ from their dataset $\mathcal{D}_A$, where $|\mathcal{I}_{\text{poison}}| = \lfloor \rho \cdot N_A \rfloor$ and $\rho$ is the poisoning ratio. This selection is used to construct a \emph{dynamically} poisoned dataset $\mathcal{D}_A^{\star(t)}$ for each training epoch $t$.

The poisoning is dynamic because the trigger is not applied statically. Instead, during each epoch, a random data augmentation function $\mathcal{A}(\cdot)$ is applied to the trigger $\boldsymbol{\eta}^\star$ before embedding. For each sample index $i \in \{1, \dots, N_A\}$, the corresponding pair $(\boldsymbol{x}'_i, y'_i)$ in the dynamic dataset is defined as:  
\begin{equation}
\label{eq:dynamic_poison}
(\boldsymbol{x}'_i, y'_i) = 
\begin{cases} 
      \big(\Pi_{\mathcal{C}}(\boldsymbol{x}_i + \mathcal{A}(\boldsymbol{\eta}^\star)),\ y_t\big) & \text{if } i \in \mathcal{I}_{\text{poison}} \\
      (\boldsymbol{x}_i,\ y_i) & \text{if } i \notin \mathcal{I}_{\text{poison}}
\end{cases}
\end{equation}

This dynamic application helps prevent the model from overfitting to the trigger during fine-tuning, thereby enhancing the backdoor's robustness. 

To select the index set $\mathcal{I}_{\text{poison}}$, we investigate two distinct strategies below.

\subsubsection{Random Selection}
As a baseline, this strategy selects $\mathcal{I}_{\text{poison}}$ by sampling $\lfloor \rho \cdot N_A \rfloor$ indices uniformly at random, without replacement, from the attacker's dataset $\mathcal{D}_A$. This method serves as a standard benchmark in data poisoning literature.

\subsubsection{Forgetting-based Selection}
In the empirical study by Toneva et al.~\cite{toneva2018empirical}, a ``forgetting event'' refers to a case where a training sample that was correctly classified at one training step is misclassified at a subsequent step. Samples that never experience such events are termed ``unforgettable.'' Their study reveals that these samples are typically ``easy'' or redundant, and a significant portion can be removed without compromising generalization.

Motivated by the observation that the distilled model's performance is influenced by the teacher's decision boundaries~\cite{song2023exploring}, we employ forgetting-based selection to control how poisoned samples perturb these boundaries. As noted in~\cite{toneva2018empirical}, samples with higher forgetting frequencies contribute more to the decision boundary, whereas those with fewer forgetting events tend to have limited influence on it. We therefore design both forgettable-based and unforgettable-based selection strategies to examine how varying sample influence on the decision boundary affects the trade-off between backdoor strength and benign accuracy.

Specifically, we record the number of forgetting events for each sample in $\mathcal{D}_A$ during the training phase of the benign model $f_{\theta_T}$. All samples are then sorted by their total forgetting counts in ascending (unforgettable-first) or descending (forgettable-first) order. The poisoning set $\mathcal{I}_{\text{poison}}$ is constructed by selecting the first $\lfloor \rho \cdot N_A \rfloor$ indices from this sorted list.

\subsection{Backdoor Injection via Fine-tuning}
\label{subsec:backdoor_injection_via_fine_tuning}

The attacker injects the backdoor by fine-tuning the benign teacher model $f_{\theta_T}$ on the generated dynamically poisoned dataset $\mathcal{D}_A^{\star(t)}$ using a \emph{reduced learning rate}, yielding the final malicious teacher $f_{\theta_T^\star}$.

To ensure the backdoor is encoded in the feature representations rather than a superficial logit-level mapping, the fine-tuning process is carefully constrained. We decompose the teacher model as $f_{\theta_T}(\boldsymbol{x}) = c_{\theta_{T, \mathrm{cls}}}(h_{\theta_{T, \mathrm{ft}}}(\boldsymbol{x}))$. During fine-tuning, the attacker \textbf{freezes} the classification layer parameters $\theta_{T, \mathrm{cls}}$. Consequently, gradients propagate only to the feature-related parameters $\theta_{T, \mathrm{ft}}$, encouraging the model to gradually embed an association between the trigger and the target class in feature space.

The fine-tuning learning rate $\eta_{\mathrm{ft}}$ is set \emph{significantly smaller} than the original training rate---typically two orders of magnitude smaller (i.e., $\eta_{\mathrm{ft}}\!=\!0.01\times\eta_{\mathrm{train}}$). This small step size is critical because a larger learning rate would easily cause the model to overfit to the trigger pattern, whereas a reduced rate promotes subtle but effective integration of the trigger–target association into the existing feature geometry, thereby effectively coupling the backdoor with the benign task. In addition, we employed a learning-rate scheduler to gradually anneal $\eta_{\mathrm{ft}}$ during fine-tuning, further stabilizing adaptation.

For each epoch $t$, only the feature extractor parameters $\theta_{T, \mathrm{ft}}$ are updated by minimizing the standard cross-entropy loss $\mathcal{L}_{\mathrm{CE}}$:
\begin{equation}
\label{eq:finetune_objective}
\min_{\theta_{T, \mathrm{ft}}} \mathbb{E}_{(\boldsymbol{x}', y') \in \mathcal{D}_A^{\star(t)}} \left[ \mathcal{L}_{\mathrm{CE}}\left(c_{\theta_{T, \mathrm{cls}}}\big(h_{\theta_{T, \mathrm{ft}}}(\boldsymbol{x}')\big), y'\right) \right],
\end{equation}

\noindent where $\theta_{T, \mathrm{cls}}$ remains fixed. The resulting parameters $\theta_{T, \mathrm{ft}}^\star$ along with $\theta_{T, \mathrm{cls}}$ constitute the malicious teacher $f_{\theta_T^\star}$.

\subsection{OSCAR: A Simplified Version of SCAR}
\label{subsec:oscar}

Concurrent work by Chen et al.~\cite{chen2025taught} introduced SCAR, a \emph{distillation-conditional backdoor attack} designed to be dormant in the teacher model but activated in the student after knowledge distillation.
However, this approach presents two notable limitations: \circled{1} its bilevel optimization is very costly (e.g., \textbf{20.75} GPU hours on an RTX 4090 for CIFAR-10, ResNet-50), hindering reproducibility; \circled{2} the attack's success is critically dependent on a pre-optimized trigger (cf. Eq. (13) in~\cite{chen2025taught}), which is constrained by an $\ell_\infty$-norm budget $\epsilon_0$ that is \emph{not evaluated}, and their own ablation study also demonstrates that replacing this trigger with a simple patch causes the attack to fail completely.

Limitation \circled{2} indicates that the trigger may function as a UAP-like pattern that is inherently transferable, and the complex optimization process might primarily trains the teacher to mask the effect of the UAP while simultaneously hindering the transfer of this masking behavior during distillation. As a result, the UAP remains effective in the student model.

To investigate this hypothesis, we implement a simplified version based on SCAR's ``w/o $\mathcal{F}_s$'' ablation study (cf. Sec. 4.3 in~\cite{chen2025taught}).
We term this baseline \textbf{OSCAR} (\underline{O}stensibly \underline{S}tealthy distillation-\underline{C}onditional b\underline{A}ckdoo\underline{R} attack). It omits the surrogate model and the complex bilevel optimization, operating in two straightforward stages:

\subsubsection{Trigger Optimization}
As proposed in~\cite{chen2025taught}, the first stage optimizes a trigger parameter $\boldsymbol{\mu}$ to be effective against \emph{both} a benign pre-trained teacher model $\hat{\mathcal{F}_{\theta_T}}: \mathcal{X}\!\to\!\mathbb{R}^C$, and its benignly distilled student $\hat{\mathcal{F}_{\theta_S}}: \mathcal{X}\!\to\!\mathbb{R}^C$.
Let the trigger injection function be $G(\boldsymbol{x}; \boldsymbol{\mu}) = \Pi_{\mathcal{C}}(\boldsymbol{x} + \boldsymbol{\mu})$.
The optimal trigger $\boldsymbol{\mu}^\star$ is found by solving:
\begin{equation}
\label{eq:oscar_trigger}
\begin{aligned}
\boldsymbol{\mu}^\star = \underset{\boldsymbol{\mu}}{\arg\min}\;
&\mathbb{E}_{(\boldsymbol{x},y)\in \mathcal{D}_A}
\Bigl[ \mathcal{L}_{\mathrm{CE}}(\hat{\mathcal{F}_{\theta_T}}(G(\boldsymbol{x};\boldsymbol{\mu})), y_t) \\
&\quad + \mathcal{L}_{\mathrm{CE}}(\hat{\mathcal{F}_{\theta_S}}(G(\boldsymbol{x};\boldsymbol{\mu})), y_t) \Bigr], \\
&\text{s.t.}\; \|\boldsymbol{\mu}\|_\infty \le \epsilon_0 .
\end{aligned}
\end{equation}

\subsubsection{Teacher Masking}
In the second stage, a new teacher model $\mathcal{F}_{\theta_T}: \mathcal{X}\!\to\!\mathbb{R}^C$ is trained to actively ``mask'' the backdoor behavior.
This masking is enforced by minimizing the following composite loss function:
\begin{equation}
\label{eq:oscar_finetune}
\begin{aligned}
\underset{\theta_T}{\min}\;
&\mathbb{E}_{(\boldsymbol{x},y)\in \mathcal{D}_A} \Bigl[ \mathcal{L}_\mathrm{CE}(\mathcal{F}_{\theta_T}(\boldsymbol{x}),y) \\
&\quad + \alpha_m \cdot \mathcal{L}_\mathrm{CE}(\mathcal{F}_{\theta_T}(G(\boldsymbol{x}; \boldsymbol{\mu}^\star)),y)\Bigr]
\end{aligned}
\end{equation}
\noindent where $\alpha_m$ controls the trade-off between benign accuracy and trigger suppression.

\section{Experiments}

\begin{table*}[!t]
\centering
\caption{Performance (\%) of ADBA and BackWeak on CIFAR-10 and ImageNet-50. \textbf{TO} denotes \emph{Trigger Only}, where triggered inputs are tested on \textbf{benign models} without any backdoor injection. Noteworthy values are highlighted in \textcolor{red}{red}.}
\label{tab:main_results}
\setlength{\tabcolsep}{4pt}
\begin{tabular}{ccccccccccccccc} 
\toprule
\multirow{3}{*}{Dataset}   & \multirow{3}{*}{KD Method} & Model$\rightarrow$ & \multicolumn{3}{c}{ResNet-34}                                                       & \multicolumn{3}{c}{MobileNet-V2}                                                    & \multicolumn{3}{c}{ShuffleNet-V2}                                                            & \multicolumn{3}{c}{DenseNet-BC-121}                                                  \\
                           &                            &                    & \multicolumn{3}{c}{(Teacher)}                                                       & \multicolumn{3}{c}{(Student A)}                                                     & \multicolumn{3}{c}{(Student B)}                                                              & \multicolumn{3}{c}{(Student C)}                                                      \\ 
\cmidrule{4-15}
                           &                            & Attack$\downarrow$ & ASR                     & BA                              & TITG                    & ASR                     & BA                              & TITG                    & ASR                              & BA                              & TITG                    & ASR                     & BA                              & TITG                     \\ 
\midrule
\multirow{12}{*}{CIFAR-10} & \multirow{4}{*}{Response}  & ADBA (TO)          & \textcolor{red}{94.41} & 90.63                           & \textcolor{red}{85.06} & \textcolor{red}{95.64} & 90.33                           & \textcolor{red}{86.40} & \textcolor{red}{\textbf{95.33}} & 90.39                           & \textcolor{red}{86.19} & \textcolor{red}{96.40} & 91.24                           & \textcolor{red}{86.90}  \\
                           &                            & ADBA               & 97.64                   & 73.75                           & 71.30                   & 91.62                   & 80.34                           & 70.77                   & 94.27                            & 80.49                           & 73.24                   & 95.63                   & 80.74                           & 74.43                    \\
                           &                            & BackWeak (TO)      & 4.89                    & 90.63                           & 3.17                    & 3.73                    & 90.33                           & 1.95                    & 3.40                             & 90.39                           & 1.54                    & 3.63                    & 91.24                           & 2.03                     \\
                           &                            & BackWeak           & \textbf{99.97}          & 87.97                           & 89.03                   & \textbf{97.50}          & 89.79                           & 87.75                   & 93.07                            & 89.88                           & 83.61                   & \textbf{99.21}          & 90.58                           & 89.34                    \\ 
\cmidrule{2-15}
                           & \multirow{4}{*}{Feature}   & ADBA (TO)          & \textcolor{red}{94.41} & 90.63                           & \textcolor{red}{85.06} & \textcolor{red}{95.77} & 87.73                           & \textcolor{red}{86.59} & \textcolor{red}{89.53}          & 87.00                           & \textcolor{red}{80.33} & \textcolor{red}{96.67} & 89.59                           & \textcolor{red}{86.88}  \\
                           &                            & ADBA               & 97.64                   & 73.75                           & 71.30                   & 92.62                   & 73.34                           & 65.89                   & \textbf{93.56}                  & 72.59                           & 65.78                   & 96.79                   & 73.49                           & 69.32                    \\
                           &                            & BackWeak (TO)      & 4.89                    & 90.63                           & 3.17                    & 4.13                    & 87.73                           & 1.92                    & 4.39                             & 87.00                           & 1.57                    & 4.54                    & 89.59                           & 2.46                     \\
                           &                            & BackWeak           & \textbf{99.97}          & 87.97                           & 89.03                   & \textbf{98.96}          & 85.56                           & 88.23                   & 89.13                            & 84.70                           & 79.21                   & \textbf{99.88}          & 88.05                           & 88.48                    \\ 
\cmidrule{2-15}
                           & \multirow{4}{*}{Relation}  & ADBA (TO)          & \textcolor{red}{94.41} & 90.63                           & \textcolor{red}{85.06} & \textcolor{red}{96.61} & 89.78                           & \textcolor{red}{87.17} & \textcolor{red}{85.36}          & 89.06                           & \textcolor{red}{76.87} & \textcolor{red}{97.04} & 89.86                           & \textcolor{red}{87.70}  \\
                           &                            & ADBA               & 97.64                   & 73.75                           & 71.30                   & 89.76                   & 75.70                           & 64.82                   & \textbf{89.80}                  & 75.22                           & 64.20                   & 90.68                   & 76.29                           & 66.06                    \\
                           &                            & BackWeak (TO)      & 4.89                    & 90.63                           & 3.17                    & 3.83                    & 89.78                           & 1.99                    & 3.38                             & 89.06                           & 1.23                    & 4.46                    & 89.86                           & 2.67                     \\
                           &                            & BackWeak           & \textbf{99.97}          & 87.97                           & 89.03                   & \textbf{97.88}          & 88.71                           & 87.90                   & 60.77                            & 88.43                           & 53.96                   & \textbf{97.82}          & 88.16                           & 87.34                    \\ 
\midrule
\multirow{12}{*}{ImageNet-50} & \multirow{4}{*}{Response}  & ADBA (TO)          & 1.18                    & 74.09                           & 5.96                    & 0.77                    & 76.29                           & 0.00                    & 1.09                             & 76.59                           & 0.34                    & 1.07                    & 75.16                           & 0.00                     \\
                           &                            & ADBA               & 90.09                   & \textcolor{red}{\underline{27.34}} & 17.63                   & 62.02                   & \textcolor{red}{\underline{51.99}} & 17.77                   & 53.77                            & \textcolor{red}{\underline{54.43}} & 13.41                   & 59.98                   & \textcolor{red}{\underline{51.67}} & 16.45                    \\
                           &                            & BackWeak (TO)      & 5.76                    & 74.09                           & 5.43                    & 2.49                    & 76.29                           & 1.90                    & 2.60                             & 76.59                           & 1.89                    & 2.35                    & 75.16                           & 1.59                     \\
                           &                            & BackWeak           & \textbf{99.92}          & 71.72                           & 94.81                   & \textbf{81.43}          & 75.72                           & 78.48                   & \textbf{69.70}                   & 75.78                           & 66.55                   & \textbf{75.87}          & 74.52                           & 72.99                    \\ 
\cmidrule{2-15}
                           & \multirow{4}{*}{Feature}   & ADBA (TO)          & 1.18                    & 74.09                           & 5.96                    & 1.05                    & 70.31                           & 0.28                    & 1.12                             & 71.24                           & 0.42                    & 1.09                    & 65.16                           & 0.12                     \\
                           &                            & ADBA               & 90.09                   & \textcolor{red}{\underline{27.34}} & 17.63                   & 93.17                   & \textcolor{red}{\underline{18.90}} & 11.01                   & 92.74                            & \textcolor{red}{\underline{19.52}} & 11.19                   & \textbf{96.01}         & \textcolor{red}{\underline{13.89}} & 8.55                     \\
                           &                            & BackWeak (TO)      & 5.76                    & 74.09                           & 5.43                    & 3.33                    & 70.31                           & 2.72                    & 3.37                             & 71.24                           & 3.10                    & 4.03                    & 65.16                           & 3.40                     \\
                           &                            & BackWeak           & \textbf{99.92}          & 71.72                           & 94.81                   & \textbf{97.12}          & 68.10                           & 92.17                   & \textbf{98.54}                   & 69.34                           & 93.56                   & 91.87                   & 62.62                           & 86.86                    \\ 
\cmidrule{2-15}
                           & \multirow{4}{*}{Relation}  & ADBA (TO)          & 1.18                    & 74.09                           & 5.96                    & 0.81                    & 75.60                           & 0.08                    & 1.05                             & 75.60                           & 0.32                    & 1.36                    & 76.05                           & 0.70                     \\
                           &                            & ADBA               & 90.09                   & \textcolor{red}{\underline{27.34}} & 17.63                   & 85.02                   & \textcolor{red}{\underline{30.58}} & 15.12                   & \textbf{80.60}                  & \textcolor{red}{\underline{29.83}} & 70.75                   & 82.24                   & \textcolor{red}{\underline{31.48}} & 13.35                    \\
                           &                            & BackWeak (TO)      & 5.76                    & 74.09                           & 5.43                    & 2.68                    & 75.60                           & 2.17                    & 2.27                             & 75.60                           & 1.61                    & 2.90                    & 76.05                           & 2.44                     \\
                           &                            & BackWeak           & \textbf{99.92}          & 71.72                           & 94.81                   & \textbf{95.84}          & 74.46                           & 91.61                   & 78.93                            & 73.81                           & 74.78                   & \textbf{94.57}          & 74.32                           & 90.36                    \\
\bottomrule
\end{tabular}
\end{table*}

\subsection{Experimental Setup}
\label{sec:exp_setup}

\noindent\mbox{\textbf{Datasets.}} We mainly evaluate our method on two benchmark image datasets: CIFAR-10~\cite{krizhevsky2009learning} and ImageNet-50~\cite{deng2009imagenet}. The ImageNet-50 dataset is constructed by randomly selecting 50 classes from the standard ImageNet dataset. To simulate a setting where the attacker and victim have different training data, we randomly split each training set into two \emph{equal-size, disjoint} subsets and use them as the attacker dataset $\mathcal{D}_A$ and the victim (distillation) dataset $\mathcal{D}_V$. Validation sets remain unchanged and are used for final evaluation. More details are provided in Appendix~A.

\vspace{0.3em}

\noindent\mbox{\textbf{Models and KD Methods.}} For teacher models, we employ high-capacity networks, including ResNet-34~\cite{he2016deep}, VGG-16~\cite{simonyan2014very}, and RegNet-Y~\cite{radosavovic2020designing}, while lightweight architectures such as MobileNet-V2~\cite{sandler2018mobilenetv2}, ShuffleNet-V2~\cite{ma2018shufflenet} and DenseNet-BC-121~\cite{huang2017densely} are used as student models. To comprehensively evaluate our method, we adopt three representative knowledge distillation (KD) techniques, each corresponding to a distinct paradigm: \emph{response-based}~\cite{hinton2015distilling}, \emph{feature-based}~\cite{chen2022knowledge}, and \emph{relation-based}~\cite{huang2022knowledge} distillation.

\vspace{0.3em}

\noindent\mbox{\textbf{Evaluation Metrics.}} We evaluate the attack effectiveness using the \emph{\textbf{Attack Success Rate (ASR)}} metric, defined as the proportion of \emph{non-target-class} samples successfully misclassified as the target class. We also report the \emph{\textbf{Benign Accuracy (BA)}}, defined as the classification accuracy on clean test samples, to measure the model's performance on benign tasks. Additionally, to quantify the trigger's intrinsic adversarial effect on \emph{benign} models, we introduce the \emph{\textbf{Trigger-Induced Target Gain (TITG)}}. TITG measures the increase in the prediction probability for the target class $y_t$ when the trigger is applied to the input---specifically, the increase over the baseline probability (e.g., $p_0$ defined in Sec.~\ref{subsubsec:weak_trigger_def}). Therefore, we can directly compare the TITG on \emph{benign models} with $\epsilon_t$ in Eq.~\ref{eq:weak_trigger} to determine whether the trigger is a weak trigger, $\epsilon_t$ is typically set to 0.1. Note that we also report TITG values computed on malicious models.  

\vspace{0.3em}

\noindent\mbox{\textbf{Baselines and Attack Settings.}} We compare \texttt{BackWeak} to the classic surrogate-based baseline ADBA~\cite{ge2021anti} (not open-source, re-implemented by us), while also utilizing OSCAR to investigate the strong adversarial nature of the SCAR~\cite{chen2025taught} trigger. Our implementations follow the original papers, with ADBA targeting the same attacker objective and tuned to achieve comparable attack effect for fair performance comparison, while OSCAR (though using a different attacker objective) is used mainly to calibrate the trigger-constraint param $\epsilon_0$ (cf. Eq. (13) in~\cite{chen2025taught}) and to probe the trigger's stealthiness and intrinsic adversarial properties. Unless specified, we use SGD (momentum 0.9, weight decay $5 \times 10^{-4}$), a cosine annealing scheduler, and a batch size of 128 for all processes. For the \texttt{BackWeak} workflow: \emph{Stage 1 (Model Training)} uses $\eta_{\mathrm{train}}=10^{-2}$ for 200 epochs. \emph{Stage 2 (Weak Trigger Generation)} sets $\delta_{\text{ASR}}=0.05$, $\lambda=1$, $\mu=1$, $\epsilon_0=8/255$, and optimizes for 200 epochs with $\eta_{\mathrm{tri}}=10^{-4}$ (CIFAR-10) or $10^{-2}$ (ImageNet-50). \emph{Stage 3 (Sample Selection and Data Poisoning)} employs random selection with $\rho=0.3$, applying random flip and crop augmentations ($\mathcal{A}$) to the trigger. \emph{Stage 4 (Backdoor Injection)} fine-tunes for 150 epochs with the last layer frozen, using $\eta_{\mathrm{ft}}=10^{-4}$. All evaluation distillations run for 200 epochs with an initial learning rate of $10^{-2}$. Specific KD settings are: $\alpha=0.5, \tau=5$ for Vanilla KD (cf. Eq.~\ref{eq:distill}); direct feature alignment for Feature-based KD (cf. Eq. (3) in \cite{chen2022knowledge}); and recommended $\alpha=1, \beta=2, \gamma=2$ for Relation-based KD (cf. Sec. 4.1 in \cite{huang2022knowledge}), with temperature $\tau$ set to 4. More details are in Appendix~A.

\subsection{Attack Experiments}
\label{sec:attack_experiments}

\noindent\mbox{\textbf{Main Results.}} We empirically evaluate BackWeak against ADBA across three student architectures and two datasets; results are shown in Table~\ref{tab:main_results}. For a fair comparison, we constrain ADBA's trigger values to $[0, \frac{1}{c}]$ with $c=31$ (cf. Sec 3.2.3 in \cite{ge2021anti}) and independently train/distill benign teacher and student models for both methods. In the table, best ASRs are \textbf{bolded} and severe BA degradations ($>20\%$) are \underline{underlined}. Overall, BackWeak consistently achieves high attack success rates while incurring minimal benign accuracy loss (typically $< 3\%$). By contrast, ADBA frequently causes catastrophic utility degradation, especially on the larger ImageNet-50 dataset where BA often drops by more than $20\%$. Critically, the ``Trigger Only'' (TO) experiments on benign models reveal important differences in trigger properties: on CIFAR-10, the ADBA trigger exhibits a strong intrinsic adversarial effect (e.g., $>70\%$ TITG), functioning as a UAP that achieves high ASR even without backdoor injection, meaning that, \emph{ADBA's process resembles constructing UAPs rather than implanting a true backdoor} and the resulting triggers cannot be considered ``weak'' according to Eq.~\ref{eq:weak_trigger}; on ImageNet-50, where this UAP effect is weak due to larger image size, the ADBA process severely damages the model (low BA) to perform the backdoor injection. Conversely, BackWeak's trigger consistently produce a low TITG ($< 10\%$) on benign models, satisfying our definition of a ``weak trigger''. This confirms that BackWeak successfully implants a genuine, transferable backdoor using a weak trigger, rather than strongly relying on the trigger's adversarial property. Additional results on VGG-16 and RegNet are provided in Appendix~B.

\begin{table}[htbp]
\centering
\caption{Performance (\%) of OSCAR with varying $\epsilon_0$ constraints. Non-weak TITGs ($>10\%$) are marked in \textcolor{red}{red}.}
\label{tab:oscar_epsilon}
\setlength{\tabcolsep}{3pt}
\begin{tabular}{ccccccc} 
\toprule
\multirow{2}{*}{} & Teacher                 & \multicolumn{2}{c}{Teacher} & Student         & Student         & Student          \\
                  & (Benign)                & \multicolumn{2}{c}{(OSCAR)} & (Response)      & (Feature)       & (Relation)       \\ 
\cmidrule{2-7}
$\epsilon_0$      & TITG                    & ASR$\downarrow$ & TITG      & ASR$\uparrow$   & ASR$\uparrow$   & ASR$\uparrow$    \\ 
\midrule
8/255             & \textcolor{red}{14.41} & 1.30            & 0.00      & 4.49            & 4.73            & 4.87             \\
16/255            & \textcolor{red}{69.68} & 1.40            & 0.13      & 32.52           & 34.04           & 29.92            \\
24/255            & \textcolor{red}{87.95} & 1.54            & 0.02      & \textbf{61.16} & \textbf{67.08} & \textbf{68.50}  \\
32/255            & \textcolor{red}{89.80} & 1.51            & 0.17      & \textbf{87.11} & \textbf{84.24} & \textbf{88.12}  \\
\bottomrule
\end{tabular}
\end{table}

\noindent\mbox{\textbf{Effect of Trigger Strength on OSCAR.}} We next investigate how trigger strength, controlled by the $\ell_\infty$ constraint $\epsilon_0$, influences the effectiveness of the SCAR attack. Using OSCAR, a simplified SCAR baseline, we conduct experiments on CIFAR-10 with a benign and an OSCAR ResNet-34 teacher, and MobileNet-V2 students distilled from the latter. The SCAR objective is to keep the trigger inactive on the teacher while activating it on the student. As shown in Table~\ref{tab:oscar_epsilon}, noticeable attack success (student ASR $\ge50\%$, bolded) emerges only when $\epsilon_0\ge24/255$, at which point the trigger already exhibited 87.95\% TITG on the benign teacher---behaving as a strong UAP rather than a weak trigger. Even with $\epsilon_0=8/255$, the TITG of 14.41\% failed to meet our weak-trigger definition (Eq.~\ref{eq:weak_trigger}). When substituting OSCAR's trigger with BackWeak's weak trigger, the attack failed entirely, with student ASRs of 2.21\%, 2.20\%, and 2.33\% distilled by the response-based, feature-based, and relation-based KD variants, respectively. This confirms that SCAR's apparent success stems from masking a strong UAP at the teacher model rather than implanting a truly transferable backdoor.

\begin{table}[htbp]
\centering
\caption{Performance (\%) of ADBA under different $\mu$ settings. Non-weak TITGs are marked in \textcolor{red}{red}.}
\label{tab:adba_mu}
\setlength{\tabcolsep}{3pt}
\begin{tabular}{ccccccc} 
\toprule
\multirow{2}{*}{} & Teacher                 & \multicolumn{2}{c}{Teacher} & Student    & Student   & Student     \\
                  & (Benign)                & \multicolumn{2}{c}{(ADBA)}  & (Response) & (Feature) & (Relation)  \\ 
\cmidrule{2-7}
$\mu$             & TITG                    & ASR   & TITG                & ASR        & ASR       & ASR         \\ 
\midrule
0.1               & \textcolor{red}{85.06} & 97.64 & 71.30               & 91.62      & 92.62     & 89.77       \\
0.5               & \textcolor{red}{71.38} & 93.08 & 62.63               & 86.18      & 88.39     & 87.47       \\
1.0               & \textcolor{red}{55.28} & 90.77 & 52.26               & 76.16      & 82.92     & 79.91       \\
10.0              & 0.00                    & 8.09  & 0.01                & 6.71       & 19.00     & 11.43       \\
\bottomrule
\end{tabular}
\end{table}

\noindent\mbox{\textbf{Effect of Trigger Strength on ADBA.}} We further evaluate the impact of the hyperparameter $\mu$ in ADBA, which weights the $\ell_2$-norm constraint on the trigger mask, while keeping $c=31$ fixed. Experiments were conducted on CIFAR-10 using a benign and an ADBA ResNet-34 teacher, along with MobileNet-V2 students distilled from the latter. As shown in Table~\ref{tab:adba_mu}, increasing $\mu$ progressively weakens the adversarial strength of the trigger, as reflected by the decreasing TITG on the benign teacher. Correspondingly, the attack success rate (ASR) on the distilled students drops markedly.  Notably, when $\mu=10$, the trigger exhibits no adversarial effect (0.00 TITG) on the benign model, and the attack nearly fails (e.g., 6.71\% ASR under response-based KD). This clearly demonstrates that ADBA's effectiveness is highly dependent on the adversarial nature of its UAP-like trigger, rather than the simulated distillation process itself. 

\begin{table}[htbp]
\centering
\caption{Performance (\%) of BackWeak using a patch-based trigger.}
\label{tab:backweak_patch}
\setlength{\tabcolsep}{4pt}
\begin{tabular}{ccccccc} 
\toprule
\multirow{2}{*}{} & \multicolumn{2}{c}{Teacher}  & \multicolumn{2}{c}{Teacher}    & \multicolumn{2}{c}{Student}     \\
                  & \multicolumn{2}{c}{(Benign)} & \multicolumn{2}{c}{(BackWeak)} & \multicolumn{2}{c}{(BackWeak)}  \\ 
\cmidrule{2-7}
KD Method         & TITG & BA                    & ASR   & BA                     & ASR   & BA                      \\ 
\midrule
Response          & 0.59 & 90.63                 & 98.50 & 87.96                  & 75.26 & 90.12                   \\
Feature           & 0.59 & 90.63                 & 98.50 & 87.96                  & 87.49 & 83.47                   \\
Relation          & 0.59 & 90.63                 & 98.50 & 87.96                  & 91.42 & 88.67                   \\
\bottomrule
\end{tabular}
\end{table}

\noindent\mbox{\textbf{Adopting a Patch-based Trigger.}} We investigate whether a simple, non-optimized patch trigger (e.g., a 5$\times$5 white patch) can serve as a weak trigger within the BackWeak workflow. Such patch-based triggers, originally introduced by Gu~et~al.~\cite{gu2017badnets}, are generally semantically irrelevant to the task and typically treated as noise by benign models. As shown in Table~\ref{tab:backweak_patch}, experiments on CIFAR-10 with a ResNet-34 teacher and MobileNet-V2 students confirmed this intuition: the patch showed negligible influence (low TITG) on benign models but became highly effective after BackWeak fine-tuning (high ASR). Surprisingly, although slightly inferior to the original BackWeak in ASR, the backdoor still transferred to students across various distillation methods without a significant drop in benign accuracy. These results indicate that even a simple patch, when paired with BackWeak's low-learning-rate fine-tuning, is sufficient to implant a transferable backdoor.

\begin{figure*}[!t]
\centering
\captionsetup{font=normalsize}
\includegraphics[width=0.9\textwidth]{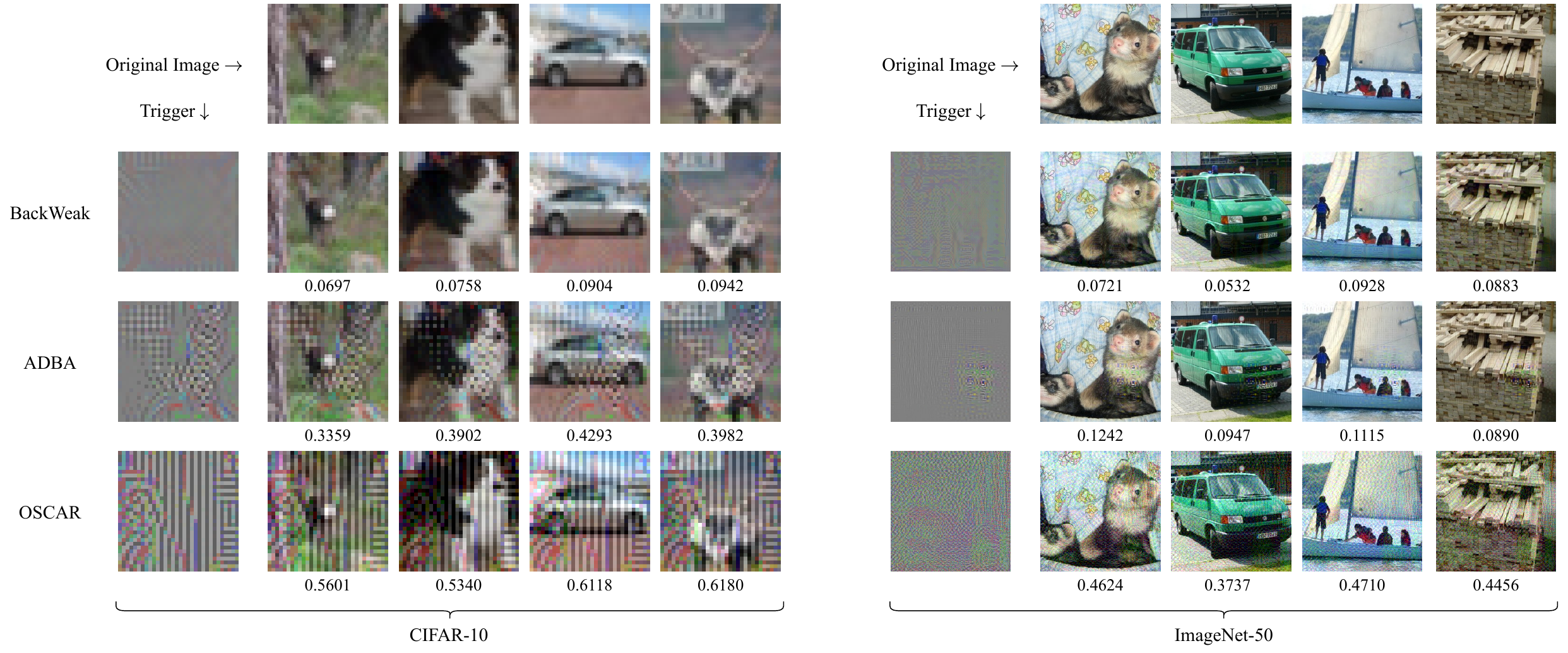}
\caption{Visualization of the trigger and its application results, with LPIPS values (computed against the original images) displayed below each image obtained by applying the trigger.}
\label{fig:trigger_visualization} 
\end{figure*}

\subsection{Perceptibility of Triggers}

Figure~\ref{fig:trigger_visualization} presents a visual and quantitative comparison of triggers generated by BackWeak, ADBA, and OSCAR under parameter settings tuned for high ASR. Specifically, BackWeak uses $\epsilon_0=8/255$, ADBA follows \cite{ge2021anti} with $\mu=0.1$ and $c=1$, and OSCAR uses $\epsilon_0=32/255$ (cf. Sec.~\ref{sec:attack_experiments}). We measure \emph{Learned Perceptual Image Patch Similarity} (LPIPS)~\cite{zhang2018perceptual} between original and triggered images, where a smaller LPIPS value indicates that the triggered image is more similar to the original one. Both qualitative (visual) inspection and quantitative LPIPS results indicate that the BackWeak trigger is stealthier. In contrast, ADBA and OSCAR require strong, perceptible perturbations to achieve high ASR, which compromises their stealthiness. These findings demonstrate that BackWeak can implant a transferable backdoor using an imperceptible trigger, whereas the baselines rely on visually apparent, strong UAP-like triggers.  

\subsection{Parameter Analysis and Ablation Studies}
\label{sec:param_analysis_ablation_studies}

In this section, we analyze the impact of key parameters in BackWeak and present ablation studies to assess the role and necessity of several components. Unless specified, we employ the \emph{unforgettable-first} strategy in Stage 3 (cf. Sec~\ref{subsec:sample_selection_and_data_poisoning}) throughout this section to ensure consistency in sample selection, and we adopt ResNet-34 as the teacher model architecture and MobileNet-V2 as the student model architecture.

\subsubsection{Attacker Side Analysis} 
We first analyze the key factors on the attacker's side that affect attack effectiveness.  

\begin{table*}[!t]
\centering
\caption{Performance (\%) of BackWeak under different trigger $\ell_\infty$ constraints $\epsilon_0$.}
\label{tab:backweak_param_eps0}
\setlength{\tabcolsep}{5pt}
\begin{tabular}{ccccccccccccc} 
\toprule
\multirow{3}{*}{Dataset}     & \multirow{2}{*}{} & \multicolumn{3}{c}{Teacher}  & \multicolumn{2}{c}{Teacher}    & \multicolumn{2}{c}{Student}    & \multicolumn{2}{c}{Student}   & \multicolumn{2}{c}{Student}     \\
                             &                   & \multicolumn{3}{c}{(Benign)} & \multicolumn{2}{c}{(BackWeak)} & \multicolumn{2}{c}{(Response)} & \multicolumn{2}{c}{(Feature)} & \multicolumn{2}{c}{(Relation)}  \\ 
\cmidrule{3-13}
                             & $\epsilon_0$      & ASR  & BA    & TITG          & ASR    & BA                    & ASR   & BA                     & ASR   & BA                    & ASR   & BA                      \\ 
\midrule
\multirow{3}{*}{CIFAR-10}    & 8/255             & 4.89 & 90.63 & 3.17          & 99.87  & 87.17                 & 97.43 & 89.75                  & 98.67 & 85.14                 & 98.53 & 88.33                   \\
                             & 16/255            & 5.69 & 90.63 & 4.07          & 99.97  & 88.29                 & 99.53 & 89.70                  & 99.28 & 85.21                 & 99.60 & 88.81                   \\
                             & 32/255            & 6.20 & 90.63 & 4.61          & 100.00 & 89.07                 & 99.47 & 89.74                  & 98.00 & 85.77                 & 98.44 & 89.45                   \\ 
\midrule
\multirow{3}{*}{ImageNet-50} & 8/255             & 5.76 & 74.09 & 5.43          & 99.78  & 70.87                 & 72.61 & 75.70                  & 93.45 & 67.39                 & 89.09 & 74.15                   \\
                             & 16/255            & 7.11 & 74.09 & 6.72          & 99.98  & 71.92                 & 42.80 & 75.83                  & 75.32 & 68.58                 & 88.00 & 74.76                   \\
                             & 32/255            & 5.80 & 74.09 & 5.41          & 99.98  & 72.16                 & 67.38 & 75.95                  & 89.29 & 68.60                 & 94.14 & 74.80                   \\
\bottomrule
\end{tabular}
\end{table*}

\vspace{0.3em}

\noindent\mbox{\textbf{Trigger Constraint $\epsilon_0$.}} The hyperparameter $\epsilon_0$ controls the trigger's perceptibility via the $\ell_\infty$-norm constraint (cf. Eq.~\ref{eq:weak_trigger_opt}); a larger value makes the trigger more conspicuous, thereby making the trigger feature-wise more separable from clean images. We evaluate BackWeak's effectiveness on CIFAR-10 and ImageNet-50 with $\epsilon_0$ values of 8/255, 16/255, and 32/255. As shown in Table~\ref{tab:backweak_param_eps0}, while a larger $\epsilon_0$ increases perceptibility, the BackWeak generation process successfully maintains low trigger adversariality (TITG $< 7\%$ on the benign teacher). For the BackWeak teacher, the increased separability facilitates learning the trigger-target association, leading to higher ASR. Such improved separability also reduces interference with clean image features, generally improving BA in both teacher and student models. However, this decoupling of trigger features from the clean features can destabilize the backdoor embedded in ``dark knowledge'' transferred during distillation with a clean dataset, potentially causing ASR fluctuations in the student, as observed in the ImageNet-50 response-based distillation results. Despite this, student models generally maintain a high ASR. Note that $\epsilon_0$ is typically tuned together with $\mu$ and $\delta_{\text{ASR}}$; varying $\epsilon_0$ alone primarily enhances the trigger's separability from image features rather than increasing its intrinsic adversariality.

\begin{table*}[!t]
\centering
\caption{Performance (\%) of BackWeak under different trigger adversariality constraints $\mu$ and $\delta_{\text{ASR}}$. Non-weak TITGs ($>10\%$) are marked in \textcolor{red}{red}.}
\label{tab:backweak_param_mu_budget}
\begin{tabular}{cccccccccccccc} 
\toprule
\multirow{3}{*}{Dataset}     &       & \multirow{2}{*}{}   & \multicolumn{3}{c}{Teacher}             & \multicolumn{2}{c}{Teacher}    & \multicolumn{2}{c}{Student}    & \multicolumn{2}{c}{Student}   & \multicolumn{2}{c}{Student}     \\
                             &       &                     & \multicolumn{3}{c}{(Benign)}            & \multicolumn{2}{c}{(BackWeak)} & \multicolumn{2}{c}{(Response)} & \multicolumn{2}{c}{(Feature)} & \multicolumn{2}{c}{(Relation)}  \\ 
\cmidrule{4-14}
                             & $\mu$ & $\delta_\text{ASR}$ & ASR   & BA    & TITG                    & ASR    & BA                    & ASR   & BA                     & ASR   & BA                    & ASR   & BA                      \\ 
\midrule
\multirow{3}{*}{CIFAR-10}    & 1.0   & 0.05                & 5.69  & 90.63 & 4.07                    & 99.97  & 88.29                 & 99.53 & 89.70                  & 99.20 & 85.21                 & 99.60 & 88.81                   \\
                             & 2.0   & 0.10                & 13.41 & 90.63 & \textcolor{red}{11.10} & 100.00 & 89.54                 & 99.73 & 90.11                  & 99.53 & 86.62                 & 99.64 & 89.55                   \\
                             & 5.0   & 0.20                & 16.86 & 90.63 & \textcolor{red}{12.64} & 100.00 & 89.96                 & 99.93 & 90.37                  & 99.94 & 87.25                 & 99.09 & 89.55                   \\ 
\midrule
\multirow{3}{*}{ImageNet-50} & 1.0   & 0.05                & 7.12  & 74.09 & 6.72                    & 99.98  & 71.92                 & 42.80 & 75.83                  & 75.32 & 68.58                 & 88.00 & 74.76                   \\
                             & 2.0   & 0.10                & 11.33 & 74.09 & \textcolor{red}{10.81} & 100.00 & 72.24                 & 69.04 & 76.03                  & 82.24 & 68.50                 & 85.10 & 74.70                   \\
                             & 5.0   & 0.20                & 21.86 & 74.09 & \textcolor{red}{21.26} & 99.98  & 72.67                 & 88.77 & 76.51                  & 95.48 & 69.28                 & 96.49 & 75.58                   \\
\bottomrule
\end{tabular}
\end{table*}

\begin{table}[htbp]
\centering
\caption{Performance (\%) of BackWeak under different poisoning ratios $\rho$.}
\label{tab:backweak_rho}
\setlength{\tabcolsep}{4pt}
\begin{tabular}{ccccccccc} 
\toprule
       & \multicolumn{2}{c}{Teacher}    & \multicolumn{2}{c}{Student}    & \multicolumn{2}{c}{Student}   & \multicolumn{2}{c}{Student}     \\
       & \multicolumn{2}{c}{(BackWeak)} & \multicolumn{2}{c}{(Response)} & \multicolumn{2}{c}{(Feature)} & \multicolumn{2}{c}{(Relation)}  \\ 
\cmidrule{2-9}
$\rho$ & ASR   & BA                     & ASR   & BA                     & ASR   & BA                    & ASR   & BA                      \\ 
\midrule
0.1    & 99.42 & 88.71                  & 95.97 & 90.09                  & 92.92 & 86.42                 & 95.77 & 89.41                   \\
0.3    & 99.87 & 87.17                  & 97.43 & 89.75                  & 98.67 & 85.14                 & 98.53 & 88.33                   \\
0.5    & 99.93 & 83.97                  & 98.44 & 89.05                  & 99.63 & 81.85                 & 99.37 & 86.40                   \\
\bottomrule
\end{tabular}
\end{table}

\vspace{0.3em}

\noindent\mbox{\textbf{Trigger Constraints $\mu, \delta_{\text{ASR}}$.}} The hyperparameters $\mu$ and $\delta_{\text{ASR}}$ jointly constrain the trigger's intrinsic adversariality (cf. Eqs.~\ref{eq:margin_loss}, \ref{eq:weak_trigger_opt}); relaxing these constraints (i.e., using larger values) allows the generation of a more potent trigger. We evaluate this effect on CIFAR-10 and ImageNet-50, increasing $(\mu, \delta_{\text{ASR}})$ from $(1.0, 0.05)$ to $(5.0, 0.20)$. As shown in Table~\ref{tab:backweak_param_mu_budget}, relaxing these constraints significantly boosts the trigger's inherent adversarial strength, evidenced by the rising TITG on the benign teacher (e.g., from 6.72\% to 21.26\% on ImageNet-50). This stronger signal, in turn, substantially improves the ASRs in the distilled student models, particularly on the more complex ImageNet-50 dataset. Interestingly, relaxing these constraints also leads to a slight improvement in BA on both datasets. We believe this is because a stronger trigger is inherently closer to the target class in the model's feature space, making the triggered samples lie nearer to the decision boundary. As a result, the model can learn the backdoor association with less distortion to its decision boundaries. This effect carries over to the student models as well: since the teacher's backdoored decision boundary is easier to mimic, the students can maintain higher BA after distillation.

\begin{table}[htbp]
\centering
\caption{Performance (\%) of BackWeak under different sample selection strategies, where R, F, and U denote the Random, Forgettable-first, and Unforgettable-first strategies, respectively.}
\label{tab:backweak_sample_sel}
\setlength{\tabcolsep}{3pt}
\begin{tabular}{ccccccccc} 
\toprule
         & \multicolumn{2}{c}{Teacher}    & \multicolumn{2}{c}{Student}    & \multicolumn{2}{c}{Student}   & \multicolumn{2}{c}{Student}     \\
         & \multicolumn{2}{c}{(BackWeak)} & \multicolumn{2}{c}{(Response)} & \multicolumn{2}{c}{(Feature)} & \multicolumn{2}{c}{(Relation)}  \\ 
\cmidrule{2-9}
Strategy & ASR   & BA                     & ASR   & BA                     & ASR   & BA                    & ASR   & BA                      \\ 
\midrule
R        & 99.92 & 71.72                  & 81.43 & 75.72                  & 97.12 & 68.10                 & 95.84 & 74.46                   \\
F        & 97.71 & 71.28                  & 79.14 & 75.42                  & 92.70 & 67.91                 & 88.54 & 74.19                   \\
U        & 99.78 & 70.87                  & 72.61 & 75.70                  & 93.45 & 67.39                 & 89.09 & 74.15                   \\
\bottomrule
\end{tabular}
\end{table}

\vspace{0.3em}

\noindent\mbox{\textbf{Poisoning Ratio $\rho$.}} The poisoning ratio $\rho$ controls the strength of the backdoor signal during the teacher model's fine-tuning (cf. Sec.~\ref{subsec:sample_selection_and_data_poisoning}), where a larger $\rho$ is generally expected to improve attack effectiveness. We validate the impact of $\rho$ on CIFAR-10, setting its value to 0.1, 0.3, and 0.5. As shown in Table~\ref{tab:backweak_rho}, increasing $\rho$ amplifies the backdoor signal injected during fine-tuning, which in turn enhances the signal transferred to the student. Consequently, ASR improves for both the teacher and the distilled student models. However, this comes at the cost of BA, which degrades in both models. A higher poisoning ratio introduces more severe contamination of the data distribution, causing the model's learned representations to deviate from the primary task. Thus, $\rho$ serves as a critical parameter for balancing attack effectiveness and model utility.  

\vspace{0.3em}

\noindent\mbox{\textbf{Sample Selection.}} We analyze the impact of the poisoning sample selection strategy (cf. Sec.~\ref{subsec:sample_selection_and_data_poisoning}) on the more challenging ImageNet-50 dataset, comparing random (R), forgettable-first (F), and unforgettable-first (U) selection strategies. As shown in Table~\ref{tab:backweak_sample_sel}, the R strategy surprisingly provides the most effective and stable balance between ASR and BA across distillation methods. Under the U strategy, triggers are applied to samples that are least forgotten during benign training; because these samples typically lie far from the decision boundary, fine-tuning on them encourages the model to unduly distort the feature geometry, yielding slightly higher ASR than F but at the cost of a drop in BA (noting that the student may still exhibit higher BA due to its supervised loss during distillation). Conversely, the F strategy applies triggers to highly forgotten samples, which generally lie on or near the decision boundary and may even include noisy instances; learning from such samples induces substantial loss fluctuations. Given that BackWeak employs ``weak'' triggers combined with random augmentations $\mathcal{A}$ and uses a small fine-tuning learning rate, these fluctuations may hinder the formation of a stable trigger–target association in feature space, resulting in lower ASRs. Overall, random selection already provides a sufficiently effective trade-off between ASR and BA, and thus serves as a practical default strategy for BackWeak.

\begin{table}[htbp]
  \centering
  \caption{Performance (\%) of BackWeak under different fine-tuning learning rates $\eta_\mathrm{ft}$.}
    \label{tab:backweak_eta_ft}
    \setlength{\tabcolsep}{3pt}
    \begin{tabular}{ccccccccc}
    \toprule
      & \multicolumn{2}{c}{Teacher} & \multicolumn{2}{c}{Student} & \multicolumn{2}{c}{Student} & \multicolumn{2}{c}{Student} \\
      & \multicolumn{2}{c}{(BackWeak)} & \multicolumn{2}{c}{(Response)} & \multicolumn{2}{c}{(Feature)} & \multicolumn{2}{c}{(Relation)} \\
\cmidrule{2-9}    $\eta_\mathrm{ft}$ & ASR & BA & ASR & BA & ASR & BA & ASR & BA \\
    \midrule
    $10^{-2}$ & 100.00  & 91.17  & 3.64  & 90.74  & 4.06  & 84.19  & 4.33  & 89.97  \\
    $10^{-3}$ & 100.00  & 89.52  & 33.52  & 90.39  & 13.43  & 86.83  & 7.89  & 89.53  \\
    $10^{-4}$ & 99.87  & 87.17  & 97.43  & 89.75  & 98.67  & 85.14  & 98.53  & 88.33  \\
    $10^{-5}$ & 99.25  & 82.48  & 91.26  & 88.99  & 98.54  & 80.41  & 98.86  & 84.77  \\
    \bottomrule
    \end{tabular}
  \label{tab:addlabel}
\end{table}

\vspace{0.3em}

\noindent\mbox{\textbf{Learning Rate $\eta_\mathrm{ft}$.}} The fine-tuning learning rate $\eta_\mathrm{ft}$ (cf. Sec.~\ref{subsec:backdoor_injection_via_fine_tuning}) is a critical component determining the backdoor's transferability. We evaluate its impact on CIFAR-10 with values from $10^{-2}$ to $10^{-5}$, presented in Table~\ref{tab:backweak_eta_ft}. When $\eta_\mathrm{ft}$ is relatively high ($10^{-2}$ or $10^{-3}$), the teacher model achieves perfect ASR (100.00\%) but the backdoor fails to transfer during distillation, resulting in negligible student ASRs (e.g., 3.64\% for response-based KD). A significantly smaller rate, $10^{-4}$---two orders of magnitude below $\eta_{\mathrm{train}}$---is essential for effective transfer, yielding high student ASRs (e.g., $>97\%$) across all KD methods. The results also highlight a trade-off, as BA slightly degrades when $\eta_\mathrm{ft}$ decreases. We suppose that a larger $\eta_\mathrm{ft}$ drives the model to rapidly overfit the trigger through aggressive parameter updates, creating a sharp, localized decision boundary distortion that decouples the clean and triggered behaviors. Consequently, the backdoor is not encoded in the ``dark knowledge'' of clean-sample logits. In contrast, a small $\eta_\mathrm{ft}$ forces a gradual adaptation, promoting a smoother integration of the trigger-target association into the model's feature geometry. This allows the bias to be embedded within the teacher's outputs even for benign samples, enabling successful transfer via distillation.

\begin{table*}[!t]
\centering
\caption{Performance (\%) of BackWeak under different fine-tuning optimizers.} 
\label{tab:backweak_fine_tune_optimizer}
\setlength{\tabcolsep}{6pt}
\begin{tabular}{cccccccccccc} 
\toprule
\multirow{3}{*}{Dataset}     & \multirow{2}{*}{} & \multicolumn{2}{c}{Teacher}  & \multicolumn{2}{c}{Teacher}    & \multicolumn{2}{c}{Student}    & \multicolumn{2}{c}{Student}   & \multicolumn{2}{c}{Student}     \\
                             &                   & \multicolumn{2}{c}{(Benign)} & \multicolumn{2}{c}{(BackWeak)} & \multicolumn{2}{c}{(Response)} & \multicolumn{2}{c}{(Feature)} & \multicolumn{2}{c}{(Relation)}  \\ 
\cmidrule{3-12}
                             & Optimizer         & ASR  & BA                    & ASR   & BA                     & ASR   & BA                     & ASR   & BA                    & ASR   & BA                      \\ 
\midrule
\multirow{2}{*}{CIFAR-10}    & SGD               & 4.89 & 90.63                 & 99.87 & 87.17                  & 97.43 & 89.75                  & 98.67 & 85.14                 & 98.53 & 88.33                   \\
                             & Adam              & 4.89 & 90.63                 & 99.13 & 85.70                  & 94.34 & 89.52                  & 97.98 & 83.75                 & 98.04 & 87.06                   \\ 
\midrule
\multirow{2}{*}{ImageNet-50} & SGD               & 5.76 & 74.09                 & 99.78 & 70.87                  & 72.61 & 75.70                  & 93.45 & 67.39                 & 89.09 & 74.15                   \\
                             & Adam              & 5.76 & 74.09                 & 99.47 & 69.46                  & 76.26 & 76.11                  & 95.09 & 65.64                 & 91.59 & 72.93                   \\
\bottomrule
\end{tabular}
\end{table*}

\vspace{0.3em}

\noindent\mbox{\textbf{Fine-tuning Optimizer.}} We evaluate the impact of the optimizer used during the backdoor injection fine-tuning stage (cf. Sec.~\ref{subsec:backdoor_injection_via_fine_tuning}) on CIFAR-10 and ImageNet-50, comparing SGD and Adam. Since Adam adaptively rescales parameter-wise update steps, using the same learning rate as SGD would lead to overly aggressive updates and consequently break the fine-tuning scheme required by BackWeak. To ensure comparably small effective steps, we therefore set a smaller learning rate for Adam ($\eta_\mathrm{ft}=10^{-6}$) than for SGD ($\eta_\mathrm{ft}=10^{-4}$). Other attack settings remain consistent with Sec.~\ref{sec:param_analysis_ablation_studies}. Table~\ref{tab:backweak_fine_tune_optimizer} reports the results. On the simpler CIFAR-10 dataset, SGD consistently yields higher ASRs on the distilled students, while maintaining slightly better benign accuracy. In contrast, on the more complex ImageNet-50 dataset, Adam generally achieves higher ASRs, suggesting that adaptive update scaling can sometimes help encode the trigger–target association when the data distribution is more challenging. Importantly, both optimizers preserve acceptable benign accuracy, and BackWeak remains effective regardless of the optimization choice. These results indicate that the optimizer has only a minor influence on attack performance. For BackWeak, we adopt SGD as the default optimizer, as its stability and tendency to converge to flatter minima~\cite{wu2022alignment} are conducive to forming a smoother and more transferable trigger–target association, leading to more stable backdoor injection.  

\begin{table}[htbp]
\centering
\caption{Performance (\%) of BackWeak with/without applying augmentations $\mathcal{A}$ to triggers when fine-tuning.}
\label{tab:backweak_tri_augmentation}
\setlength{\tabcolsep}{1.5pt}
\begin{tabular}{cccccccccc} 
\toprule
\multirow{3}{*}{Dataset}     & \multirow{2}{*}{} & \multicolumn{2}{c}{Teacher}    & \multicolumn{2}{c}{Student}    & \multicolumn{2}{c}{Student}   & \multicolumn{2}{c}{Student}     \\
                             &                   & \multicolumn{2}{c}{(BackWeak)} & \multicolumn{2}{c}{(Response)} & \multicolumn{2}{c}{(Feature)} & \multicolumn{2}{c}{(Relation)}  \\ 
\cmidrule{3-10}
                             & w/ $\mathcal{A}$  & ASR   & BA                     & ASR   & BA                     & ASR   & BA                    & ASR   & BA                      \\ 
\midrule
\multirow{2}{*}{CIFAR-10}    & w/                & 99.87 & 87.17                  & 97.43 & 89.75                  & 98.67 & 85.14                 & 98.53 & 88.33                   \\
                             & w/o               & 99.59 & 89.68                  & 88.60 & 90.02                  & 80.89 & 86.98                 & 87.47 & 89.65                   \\ 
\midrule
\multirow{2}{*}{ImageNet-50} & w/                & 99.78 & 70.87                  & 72.61 & 75.70                  & 93.45 & 67.39                 & 89.09 & 74.15                   \\
                             & w/o               & 99.23 & 72.52                  & 27.62 & 76.27                  & 48.93 & 69.06                 & 54.30 & 74.76                   \\
\bottomrule
\end{tabular}
\end{table}

\noindent\mbox{\textbf{Trigger Augmentation $\mathcal{A}$.}} We apply a random augmentation function $\mathcal{A}$ (cf. Eq.~\ref{eq:dynamic_poison}) to the trigger during fine-tuning to mitigate overfitting to a fixed trigger pattern. We evaluate this component on both CIFAR-10 and ImageNet-50, with results presented in Table~\ref{tab:backweak_tri_augmentation}. The findings show that omitting $\mathcal{A}$ (``w/o'') leads to a severe degradation in backdoor transferability, as reflected by the sharp drop in student ASRs, especially on ImageNet-50 (e.g., from 72.61\% to 27.62\% under response-based KD). This suggests that without augmentation, the model tends to overfit to a specific pixel pattern, creating a localized, non-transferable association. In contrast, applying $\mathcal{A}$ encourages the model to learn a more generalized and robust feature-space representation of the trigger, which, much like the effect of a low $\eta_\mathrm{ft}$, allows the backdoor to be integrated more smoothly, thereby improving its transferability.  

\begin{table}[htbp]
\centering
\caption{Performance (\%) of BackWeak with/without layer freezing when fine-tuning. T. Arch = teacher architecture; w/ Frz. = whether $c_{\theta_{T, \mathrm{cls}}}$ is frozen.}
\label{tab:backweak_layer_freeze}
\setlength{\tabcolsep}{1.5pt}
\begin{tabular}{cccccccccc} 
\toprule
\multirow{3}{*}{T. Arch}   & \multirow{2}{*}{} & \multicolumn{2}{c}{Teacher}    & \multicolumn{2}{c}{Student}    & \multicolumn{2}{c}{Student}   & \multicolumn{2}{c}{Student}     \\
                           &                   & \multicolumn{2}{c}{(BackWeak)} & \multicolumn{2}{c}{(Response)} & \multicolumn{2}{c}{(Feature)} & \multicolumn{2}{c}{(Relation)}  \\ 
\cmidrule{3-10}
                           & w/ Frz.           & ASR   & BA                     & ASR   & BA                     & ASR   & BA                    & ASR   & BA                      \\ 
\midrule
\multirow{2}{*}{ResNet-34} & w/                & 99.87 & 87.17                  & 97.43 & 89.75                  & 98.67 & 85.14                 & 98.53 & 88.33                   \\
                           & w/o               & 99.87 & 86.76                  & 96.86 & 89.85                  & 93.67 & 84.89                 & 98.41 & 88.36                   \\ 
\midrule
\multirow{2}{*}{VGG-16}    & w/                & 99.57 & 85.43                  & 97.98 & 89.45                  & 70.50 & 83.47                 & 99.33 & 87.09                   \\
                           & w/o               & 99.58 & 84.83                  & 97.60 & 89.26                  & 63.62 & 83.89                 & 99.40 & 86.79                   \\
\bottomrule
\end{tabular}
\end{table}

\vspace{0.3em}

\noindent\mbox{\textbf{Layer Freezing.}} We analyze the impact of freezing the classification layer ($c_{\theta_{T, \mathrm{cls}}}$) during the backdoor injection stage (cf. Sec.~\ref{subsec:backdoor_injection_via_fine_tuning}), a strategy designed to encourage the trigger–target association to be encoded within the feature extractor ($h_{\theta_{T, \mathrm{ft}}}$). As shown in Table~\ref{tab:backweak_layer_freeze}, experiments conducted on CIFAR-10 with both ResNet-34 and VGG-16 teachers confirm that this mechanism is particularly beneficial for feature-based distillation. Freezing the final layer markedly improves the student ASR for the feature-based method (e.g., 93.67\% $\rightarrow$ 98.67\% on ResNet-34), demonstrating that the backdoor is more effectively embedded in the feature space.

\vspace{0.4em}

\subsubsection{Victim Side Analysis} 
We next examine several critical factors on the victim's side that influence their susceptibility to attacks.  

\begin{table}[htbp]
\centering
\caption{Performance (\%) of BackWeak with varying $\alpha$ values in Vanilla KD (temperature $\tau=5$).}
\label{tab:backweak_eval_kd_alpha}
\setlength{\tabcolsep}{1pt}
\begin{tabular}{cccccccccccc} 
\toprule
\multicolumn{2}{c}{Teacher}      & \multicolumn{2}{c}{Student}        & \multicolumn{2}{c}{Student}        & \multicolumn{2}{c}{Student}        & \multicolumn{2}{c}{Student}        & \multicolumn{2}{c}{Student}         \\
\multicolumn{2}{c}{(BackWeak)}   & \multicolumn{2}{c}{($\alpha=0.1$)} & \multicolumn{2}{c}{($\alpha=0.3$)} & \multicolumn{2}{c}{($\alpha=0.5$)} & \multicolumn{2}{c}{($\alpha=0.7$)} & \multicolumn{2}{c}{($\alpha=0.9$)}  \\ 
\midrule
ASR   & BA                       & ASR   & BA                         & ASR   & BA                         & ASR   & BA                         & ASR   & BA                         & ASR   & BA                          \\ 
\midrule
99.87 & 87.17                    & 18.51 & 90.11                      & 86.14 & 90.21                      & 97.43 & 89.75                      & 98.74 & 88.70                      & 99.49 & 88.07                       \\
\bottomrule
\end{tabular}
\end{table}

\noindent\mbox{\textbf{Distillation Weight $\alpha$.}} The distillation weight $\alpha$ (see Eq.~\ref{eq:distill}) balances the student's reliance on the teacher's knowledge (distillation loss) versus the ground-truth labels (supervised loss). We evaluate this effect using Vanilla KD on CIFAR-10, with results presented in Table~\ref{tab:backweak_eval_kd_alpha}. The results indicate that a higher $\alpha$ directly correlates with increased backdoor transferability. As $\alpha$ increases from 0.1 to 0.9, the student ASR increases substantially, from 18.51\% to 99.49\%. This suggests that as the student relies more heavily on the teacher's compromised output distribution, the embedded backdoor is transferred more effectively. Meanwhile, the student's Benign Accuracy (BA) remains high and relatively stable, peaking at 90.21\% ($\alpha=0.3$) and only slightly decreasing to 88.07\% ($\alpha=0.9$), supported by the supervised loss component. Overall, the larger the $\alpha$ chosen by the victim, the more susceptible it becomes to this backdoor attack.

\begin{table}[htbp]
\centering
\caption{Performance (\%) of BackWeak with varying $\tau$ values in Vanilla KD ($\alpha=0.5$). }
\label{tab:backweak_eval_kd_tau}
\setlength{\tabcolsep}{1pt}
\begin{tabular}{cccccccccccc} 
\toprule
\multicolumn{2}{c}{Teacher}      & \multicolumn{2}{c}{Student}    & \multicolumn{2}{c}{Student}    & \multicolumn{2}{c}{Student}    & \multicolumn{2}{c}{Student}    & \multicolumn{2}{c}{Student}     \\
\multicolumn{2}{c}{(BackWeak)}   & \multicolumn{2}{c}{($\tau=1$)} & \multicolumn{2}{c}{($\tau=3$)} & \multicolumn{2}{c}{($\tau=5$)} & \multicolumn{2}{c}{($\tau=7$)} & \multicolumn{2}{c}{($\tau=9$)}  \\ 
\midrule
ASR   & BA                       & ASR   & BA                     & ASR   & BA                     & ASR   & BA                     & ASR   & BA                     & ASR   & BA                      \\ 
\midrule
99.87 & 87.17                    & 21.09 & 89.04                  & 96.00 & 89.91                  & 97.43 & 89.75                  & 97.53 & 89.39                  & 97.12 & 89.66                   \\
\bottomrule
\end{tabular}
\end{table}

\noindent\mbox{\textbf{Distillation Temperature $\tau$.}} In practice, the victim typically adjusts the temperature $\tau$ (see Eq.~\ref{eq:distill}) to soften probability distributions for effective knowledge transfer. As shown in Table~\ref{tab:backweak_eval_kd_tau} (evaluated on CIFAR-10 with Vanilla KD), the victim's choice significantly impacts susceptibility: increasing $\tau$ generally enhances backdoor transferability by exposing more latent inter-class structure, making the backdoor-induced bias more salient and easier for the student to inherit. However, when $\tau$ becomes excessively large, the logits are overly smoothed and the output distributions begin to homogenize across samples, causing the ASR to saturate or even slightly decline. Overall, moderate temperatures (e.g., $\tau=5$) make the victim most vulnerable to our attack. 

\begin{table}[htbp]
\centering
\caption{Performance (\%) of BackWeak under different victim optimizers.}
\label{tab:backweak_eval_kd_optimizer}
\setlength{\tabcolsep}{3pt}
\begin{tabular}{ccccccccc} 
\toprule
          & \multicolumn{2}{c}{Teacher}    & \multicolumn{2}{c}{Student}    & \multicolumn{2}{c}{Student}   & \multicolumn{2}{c}{Student}     \\
          & \multicolumn{2}{c}{(BackWeak)} & \multicolumn{2}{c}{(Response)} & \multicolumn{2}{c}{(Feature)} & \multicolumn{2}{c}{(Relation)}  \\ 
\cmidrule{2-9}
Optimizer & ASR   & BA                     & ASR   & BA                     & ASR   & BA                    & ASR   & BA                      \\ 
\midrule
SGD       & 99.87 & 87.17                  & 97.43 & 89.75                  & 98.67 & 85.14                 & 98.53 & 88.33                   \\
Adam      & 99.87 & 87.17                  & 96.80 & 89.55                  & 99.79 & 87.31                 & 98.29 & 88.11                   \\
\bottomrule
\end{tabular}
\end{table}

\noindent\mbox{\textbf{Distillation Optimizer.}} In practical scenarios, victims may utilize optimizers different from those used during teacher training. To evaluate the robustness of our method under such mismatched settings, we conducted experiments on CIFAR-10 where the student model was optimized using either Adam or SGD. As shown in Table~\ref{tab:backweak_eval_kd_optimizer}, the choice of optimizer exerts negligible influence on attack performance: BackWeak consistently achieves high ASRs ($>96\%$) and maintains comparable benign accuracy across all distillation paradigms, demonstrating that the transferability of the implanted backdoor is largely agnostic to the specific optimization trajectory adopted by the victim.

\begin{table}[htbp]
\centering
\caption{Performance (\%) of BackWeak on student models distilled using three KD methods across different datasets. ASR and BA are evaluated on the corresponding test sets. The BackWeak teacher is trained on CIFAR-10.}
\label{tab:backweak_eval_kd_dataset}
\setlength{\tabcolsep}{2pt}
\begin{tabular}{ccccccccc} 
\toprule
\multirow{3}{*}{Distillation Dataset} & \multicolumn{2}{c}{Teacher}    & \multicolumn{2}{c}{Student}    & \multicolumn{2}{c}{Student}   & \multicolumn{2}{c}{Student}     \\
                                      & \multicolumn{2}{c}{(BackWeak)} & \multicolumn{2}{c}{(Response)} & \multicolumn{2}{c}{(Feature)} & \multicolumn{2}{c}{(Relation)}  \\ 
\cmidrule{2-9}
                                      & ASR   & BA                     & ASR   & BA                     & ASR   & BA                    & ASR   & BA                      \\ 
\midrule
CIFAR-10                              & 99.87 & 87.17                  & 97.43 & 89.75                  & 98.67 & 85.14                 & 98.53 & 88.33                   \\
CINIC-10                              & 99.60 & 64.58                  & 97.33 & 76.40                  & 99.59 & 65.73                 & 99.57 & 66.70                   \\
\bottomrule
\end{tabular}
\end{table}

\subsection{Robustness to Data Distribution Shift}
\label{subsec:robustness_to_distribution_shift}

In Section~\ref{subsec:capability}, we assumed the attacker's dataset $\mathcal{D}_A$ and the victim's dataset $\mathcal{D}_V$ are drawn from similar distributions. We now evaluate the attack's robustness against distribution shift between the victim's $\mathcal{D}_V$ and the attacker's $\mathcal{D}_A$. Specifically, we train a BackWeak teacher on a subset of CIFAR-10 ($\mathcal{D}_A$) and distill it using two distinct victim datasets ($\mathcal{D}_V$): \circled{1} the remaining CIFAR-10 subset and \circled{2} the CINIC-10 dataset~\cite{darlow2018cinic}. We evaluate performance on the corresponding CIFAR-10 and CINIC-10 test sets.
\begin{itemize}
    \item The CINIC-10 dataset contains the same ten classes as CIFAR-10, with 9000 train and 9000 test samples per class, and is built by mixing CIFAR-10 with downsampled ImageNet images to introduce distribution shift.
\end{itemize}

As shown in Table~\ref{tab:backweak_eval_kd_dataset}, the distribution shift has a pronounced effect on benign performance: the teacher's BA drops to 64.58\% on the CINIC-10 test set, and students distilled using CINIC-10 similarly exhibit reduced BA. Interestingly, despite this BA degradation, the students distilled on CINIC-10 maintain exceptionally high ASRs, even surpassing the ASRs of students distilled on the original CIFAR-10 data. While the significant drop in BA indicates that such distillation is likely suboptimal, these results confirm that the BackWeak attack remains highly effective and transferable even when the victim employs a distillation dataset with a noticeable distribution shift.

\begin{figure*}[!t]
\centering
\subfloat[]{\includegraphics[width=2.0in]{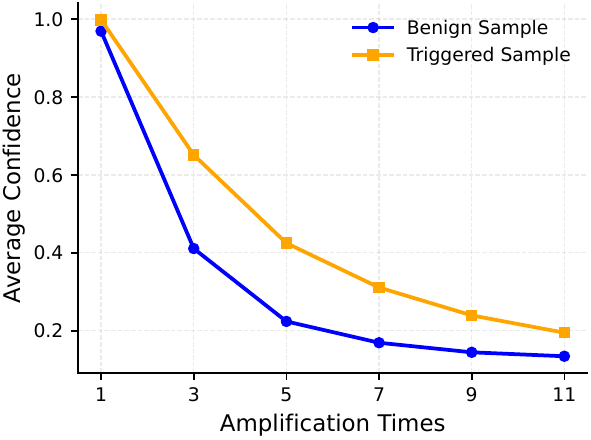}%
\label{fig:scaleup_resnet34}}
\hfil
\subfloat[]{\includegraphics[width=2.0in]{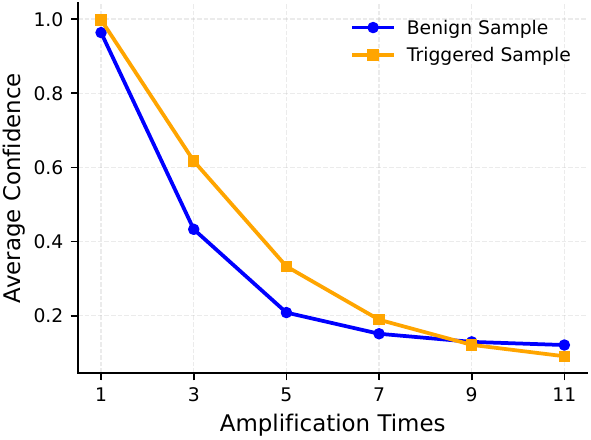}%
\label{fig:scaleup_regnet}}
\hfil
\subfloat[]{\includegraphics[width=2.0in]{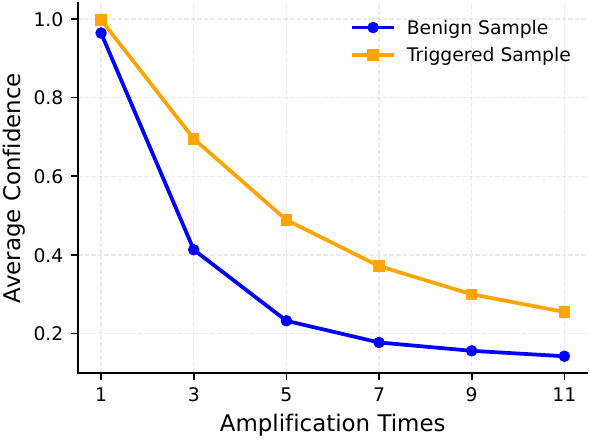}%
\label{fig:scaleup_vgg16}}
\caption{Average prediction confidence of benign and triggered samples for three BackWeak-trained models on CIFAR-10 under different pixel-level scaling factors of SCALE-UP: (a) ResNet-34, (b) RegNet-Y, and (c) VGG-16.}
\label{fig:scaleup}
\end{figure*}

\begin{table*}[!t]
  \centering
  \caption{Performance (\%) of BackWeak with and without NAD defense on teacher models. T. Arch denotes the teacher architecture.}
  \label{tab:nad_defense_backweak}%
  \setlength{\tabcolsep}{5pt}
    \begin{tabular}{ccccccccccccc}
    \toprule
      & \multicolumn{2}{c}{Teacher} & \multicolumn{2}{c}{Student} & \multicolumn{2}{c}{Teacher} & \multicolumn{2}{c}{Student} & \multicolumn{2}{c}{Teacher} & \multicolumn{2}{c}{Student} \\
      & \multicolumn{2}{c}{(Benign)} & \multicolumn{2}{c}{(Benign)} & \multicolumn{2}{c}{(w/o Defense)} & \multicolumn{2}{c}{(w/o Defense)} & \multicolumn{2}{c}{(w/ Defense)} & \multicolumn{2}{c}{(w/ Defense)} \\
\cmidrule{2-13}    T. Arch & ASR & BA & ASR & BA & ASR & BA & ASR & BA & ASR & BA & ASR & BA \\
    \midrule
    ResNet-34 & 4.89  & 90.63  & 3.73  & 90.33  & 99.97  & 87.97  & 97.51  & 89.79  & \textbf{5.76} & \underline{70.67}  & \textbf{3.34} & \underline{85.73}  \\
    VGG-16 & 4.43  & 89.36  & 3.44  & 90.04  & 99.89  & 86.14  & 99.30  & 88.90  & \textbf{5.79} & \underline{77.12}  & \textbf{3.23} & \underline{86.28}  \\
    RegNet-Y & 10.31  & 89.09  & 11.92  & 89.58  & 99.88  & 87.31  & 96.63  & 89.28  & \textbf{11.10} & \underline{63.01}  & \textbf{9.80} & \underline{83.86}  \\
    \bottomrule
    \end{tabular}%
\end{table*}%

\section{Discussion}

\subsection{Limitations under Existing Defenses}

While the proposed \texttt{BackWeak} framework is both simple and effective, we acknowledge that our study has limitations. In scenarios where downstream users are security-conscious, they may deploy backdoor defense mechanisms to detect or mitigate potential threats before initiating the distillation process. Our evaluations reveal that although \texttt{BackWeak} successfully evades certain detection-based defenses, it remains vulnerable to active backdoor mitigation strategies. Specifically, we evaluate \texttt{BackWeak} against two representative defenses:

\begin{itemize}
    \item \textbf{SCALE-UP~\cite{guo2023scale}} is a black-box, input-level backdoor detection method that identifies malicious samples by evaluating the consistency of a model's prediction confidence during a pixel-wise image amplification process. Typically, if a model's prediction confidence for an input remains highly stable and does not degrade during amplification, the input is flagged as containing a backdoor trigger. However, as illustrated in Fig.~\ref{fig:scaleup}, the model's prediction confidence for both benign and triggered samples exhibits a remarkably similar decline during the amplification process. Because the triggered samples do not display the abnormal stability characteristic of conventional backdoors, SCALE-UP fails to differentiate them from benign inputs, allowing \texttt{BackWeak} to evade detection.

    \item \textbf{Neural Attention Distillation (NAD)~\cite{li2021neural}} is a mitigation framework designed to erase backdoor triggers by aligning the intermediate-layer attention representations of the compromised network with those of a purified reference model. Operationally, this involves fine-tuning the backdoored network on a small clean dataset to obtain the reference model, followed by attention distillation to correct the intermediate feature maps. We evaluate the impact of NAD on various \texttt{BackWeak} teacher models and their distilled students using Vanilla KD on CIFAR-10. As shown in Table~\ref{tab:nad_defense_backweak}, although NAD degrades the models' Benign Accuracy (BA) (indicated by the \underline{underlined} values), it successfully and drastically reduces the Attack Success Rate (ASR) across both the teacher models and the distilled students (\textbf{bolded} values). This demonstrates that NAD effectively corrects the model's internal representations, mitigating the backdoor implanted by \texttt{BackWeak}.
\end{itemize}

The fundamental reason for this vulnerability lies in the very mechanism that enables our attack. As discussed in Sec.~\ref{subsec:knowledge_distillation}, knowledge distillation forces the student to match the teacher's predictive distribution, thereby inheriting ``dark knowledge'' such as inter-class relationships. \texttt{BackWeak} stealthily encodes the backdoor into this dark knowledge by introducing subtle shifts into the teacher's outputs for benign samples, which are then naturally transferred to the student. However, these subtle, zero-knowledge shifts are inherently fragile against significant fine-tuning and feature alignment. Mitigation strategies like NAD actively correct the model's representations, effectively erasing these delicate shifts and preventing the backdoor semantics from surviving the distillation process.

Admittedly, prior works such as ADBA~\cite{ge2021anti} and SCAR~\cite{chen2025taught} may exhibit greater resistance to such mitigation defenses. However, as our extensive analysis in Sec.~\ref{sec:attack_experiments} reveals, this apparent robustness is misleading: their triggers essentially function as Universal Adversarial Perturbations (UAPs). In essence, these methods execute strong adversarial attacks that can successfully mislead even completely benign models, rather than implanting a genuine, latent backdoor into the model's parameters. Consequently, traditional backdoor mitigation techniques, which are designed to purify poisoned parameters, fail to neutralize them.

Our work exposes the reality that many prior surrogate-based KD backdoor attacks are, to a large extent, substituting genuine backdoor injection with strong adversarial perturbations. By strictly limiting the inherent adversariality of the trigger, we demonstrate that simple fine-tuning with a small learning rate is sufficient to implant a truly transferable backdoor in the KD scenario. Nevertheless, this strict adherence to genuine backdoor principles exposes the inherent fragility of such attacks against rigorous model purification defenses. We argue that for Deep Neural Networks---especially in classification tasks---the transferability of backdoors under distillation is fundamentally delicate. Given a threat model where the attacker has zero knowledge of the victim's distillation dataset and hyperparameters (cf. Sec.~\ref{subsec:capability}), bypassing active mitigation remains a formidable challenge. Designing highly robust, transferable backdoors that maintain strict stealthiness under distillation constitutes a critical direction for future research.

Finally, to ensure a fair comparison with prior literature and to clearly demonstrate their methodological flaws, our extensive experiments were conducted within similar settings (i.e., image classification models and datasets). However, knowledge distillation is widely applied across diverse architectures, including generative models and time-series forecasting models. Exploring the dynamics of backdoor transferability and defense in these alternative domains remains an open and promising avenue for future investigation.

\subsection{Backdoor vs. Adversarial Examples: An Attacker's Perspective}

In Sec.~\ref{subsubsec:backdoor_triggers_vs_uaps}, we conceptually distinguished backdoor triggers from Universal Adversarial Perturbations (UAPs). However, empirical Attack Success Rate (ASR) metrics reveal that UAPs can achieve malicious transferability comparable to genuine backdoors in Knowledge Distillation (KD) scenarios. The primary functional divergence is that UAPs inherently compromise benign models, whereas backdoor triggers strictly activate upon models embedded with specific malicious behaviors. This raises a thought-provoking question: \textit{Given white-box access to a benign teacher model---which enables the construction of either UAPs or backdoors---why would an attacker prefer a backdoor attack? Are there scenarios where backdoors are better choices?}

We argue that despite an attacker's capability to craft UAPs, practical deployment constraints heavily favor backdoor attacks, particularly in physical-world scenarios. Constructing adversarial examples that are simultaneously highly transferable (e.g., architecture-agnostic) and physically robust (e.g., resilient to environmental variations) is notoriously challenging~\cite{wei2023visual}. Developing UAPs that satisfy both criteria for a KD pipeline is even more computationally prohibitive. Conversely, implanting a backdoor under white-box access is a significantly more tractable endeavor. As demonstrated in pioneering work like BadNets~\cite{gu2017badnets} and subsequent physical-world backdoor studies~\cite{chen2017targeted,wenger2021backdoor}, backdoor triggers serve merely as activation keys for pre-injected vulnerabilities, circumventing the need for complex, input-agnostic optimization. Consequently, backdoor triggers can be distinct, simple patterns (e.g., small patches), which inherently withstand physical environmental distortions far better than delicately crafted UAPs. Notably, physical-world backdoor attacks in KD scenarios remain an unexplored frontier, highlighting a significant gap in the field.

Furthermore, from the perspective of execution complexity, crafting cross-architecture UAPs typically necessitates surrogate student models (as seen in SCAR~\cite{chen2025taught}) to ensure transferability. In stark contrast, our \texttt{BackWeak} framework demonstrates that highly transferable KD backdoors can be easily implanted using rudimentary patches (see Table~\ref{tab:backweak_patch}) without any surrogate modeling. Given that downstream users frequently lack the awareness or resources to conduct runtime security auditing, attackers are naturally incentivized to adopt the substantially simpler and more reliable backdoor paradigm.

It is worth noting that while simple patch-based triggers are fully capable of implanting KD backdoors---and indeed exhibit the low Trigger-Induced Target Gain (TITG) required to qualify as weak triggers (Table~\ref{tab:backweak_patch})---we deliberately designed Algorithm~\ref{alg:weak_trigger_gen} to synthesize optimized weak triggers. This optimization aims to enhance visual stealthiness, rendering the trigger nearly imperceptible, while rigorously maintaining low adversariality. This stealthiness acts as a sophisticated enhancement of our framework rather than a functional prerequisite for the attack itself.

Finally, beyond practical attacker motivations, we call upon future research, particularly within the KD domain, to empirically evaluate and report the intrinsic adversariality of proposed triggers to ensure they are not merely deploying UAPs under the guise of backdoor attacks.

\section{Conclusion}

In this paper, we introduce BackWeak, a simple yet effective backdoor attack showing that a potent and transferable backdoor against knowledge distillation (KD) can be embedded by fine-tuning a benign teacher with a visually stealthy, weak trigger (i.e., a trigger with negligible intrinsic adversarial effect) under a small learning rate. This finding reveals that the computational complexity and surrogate models commonly used in existing KD backdoor attacks are not strictly necessary. Extensive experiments on CIFAR-10 and ImageNet-50 confirm that BackWeak achieves high attack success rates across diverse student models and KD methods, often surpassing prior approaches in stealthiness and efficiency. Our analysis further suggests that the apparent effectiveness of several surrogate-based methods stems less from implanting a genuine, distillation-resistant backdoor and more from the strong, UAP-like adversarial nature of their triggers. These results highlight the need for stronger scrutiny throughout the model supply chain and call on the community to more carefully assess the adversarial strength and stealthiness of triggers when studying KD backdoor attacks.

\bibliographystyle{IEEEtran}
\bibliography{refs}

\vspace{11pt}

\begin{IEEEbiographynophoto}{Shanmin Wang}
is currently working hard toward the M.S. degree in Software Engineering at Wuhan University of Technology. His research interests include backdoor attack.
\end{IEEEbiographynophoto}

\begin{IEEEbiographynophoto}{Dongdong Zhao}
(Member, IEEE) received the Ph.D. degree from University of Science and Technology of China (USTC) in 2016. He is currently an associate professor of School of Computer Science and Artificial Intelligence of Wuhan University of Technology. His research interests include: privacy-preserving deep learning, information security and biometrics.
\end{IEEEbiographynophoto}

\vfill

\appendices

\section{Implementation Details}
\label{apdx:implementation_details}

Our code, datasets, configuration files for each experiment, and the runtime environment image will be made publicly available upon publication.
Currently, our code is available at \href{https://github.com/SomeBottle/8ackW3ak}{\emph{GitHub}}.

\subsection{Details of Datasets}
\begin{enumerate}
    \item \textbf{CIFAR-10~\cite{krizhevsky2009learning}:}
    The CIFAR-10 dataset consists of 10 classes, each containing 5,000 training samples and 1,000 test samples. The entire dataset includes 50,000 training samples and 10,000 test samples. Each image has 3 channels and a resolution of $32 \times 32$.

    \item \textbf{ImageNet-50:}
    This dataset is constructed by randomly selecting 50 classes from the standard ImageNet dataset~\cite{deng2009imagenet}. Each class contains 1,000 training samples and 100 validation samples (which are disjoint from the 1,000 training samples). The dataset totals 50,000 training samples and 5,000 validation samples. Each image has 3 channels and a resolution of $224 \times 224$. We will release this subset following the ImageNet License.
\end{enumerate}

\subsection{More Details on Parameters}
\begin{enumerate}
    \item \textbf{Target Label ($y_t$):}
    We use 3 as the target label for all experiments, with one exception: for the ``\emph{Effect of Trigger Strength on OSCAR}'' analysis in Sec.~\ref{sec:attack_experiments}, we used target label 0 following the original SCAR paper~\cite{chen2025taught}.

    \item \textbf{ADBA~\cite{ge2021anti}:}
    Unless otherwise specified, we configure the ADBA teacher training process to run for 200 epochs. During the simulated distillation with a surrogate student model, the distillation KL term weight $\alpha$ is $0.5$, and the distillation temperature is $1.0$. The teacher's backdoor learning weight $\beta$ is $0.3$, and the trigger regularization term weight $\mu$ is $0.1$, which penalizes the $\ell_2$ norm of the trigger mask. Training utilized a batch size of 128. Both teacher and student models were optimized using SGD with a learning rate of $10^{-2}$ and a Cosine Annealing scheduler. The trigger is optimized using RAdam with a learning rate of $10^{-3}$.

    \item \textbf{OSCAR (Sec.~\ref{subsec:oscar}):}
    Unless otherwise specified, we also train the OSCAR teacher model for 200 epochs. The loss weight $\alpha_m$ for suppressing the backdoor trigger is set to $1.0$ (cf. Eq.~\ref{eq:oscar_finetune}), consistent with the original paper~\cite{chen2025taught}. We use a batch size of 256, an Adam optimizer with a learning rate of $10^{-4}$, and a Cosine Annealing learning rate scheduler.
\end{enumerate}

\section{Results on Additional Models}
\label{apdx:results_additional_models}

In this section, we further evaluate the effectiveness of BackWeak on the CIFAR-10 dataset using two additional teacher architectures, VGG-16~\cite{simonyan2014very} and RegNetY-1.6GF~\cite{radosavovic2020designing}, while keeping all attack settings identical to the default configuration described in Sec.~\ref{sec:exp_setup}. The corresponding results are presented in Table~\ref{tab:vgg16_results} and Table~\ref{tab:regnet_results}.

Across both teacher models, we observe that BackWeak consistently maintains high attack success rates (ASR) and high benign accuracy (BA) in the distilled students, even though the triggers used in the attack remain extremely weak. More specifically, the trigger-induced target gain (TITG) on benign teacher models generally stays below 10\%, indicating that the triggers exhibit negligible adversarial effect and satisfy our definition of weak triggers (see Eq.~\ref{eq:weak_trigger}). Despite this, BackWeak successfully implants a strong backdoor into the teacher (e.g., achieving near 100\% ASR with VGG-16), and this backdoor reliably transfers to student models distilled via response-based, feature-based, and relation-based KD.

For RegNetY-1.6GF, we observe a similar trend. Although in a few cases the TITG slightly exceeds the 10\% threshold, it still remains close to this threshold and does not resemble the strong adversarial behavior commonly seen in UAP-like triggers. Such cases can be mitigated by tightening BackWeak's adversariality constraints (e.g., using smaller $\delta_\text{ASR}$ and $\mu$ in Eqs.~\ref{eq:margin_loss}, \ref{eq:weak_trigger_opt}) during weak-trigger generation.  In contrast to triggers produced by methods such as SCAR~\cite{chen2025taught} and ADBA~\cite{ge2021anti}, which often rely on noticeably adversarial perturbations, BackWeak's triggers remain inherently weak while the implanted backdoor remains highly transferable across students.

Overall, these additional experiments further confirm BackWeak's simplicity, its use of weak triggers, and its robust transferability across different model architectures.

\begin{table*}[!t]
\centering
\caption{Performance (\%) of BackWeak on CIFAR-10 using the VGG-16 architecture. \textbf{TO} denotes \emph{Trigger Only}, where triggered inputs are tested on \textbf{benign models} without any backdoor injection.}
\label{tab:vgg16_results}
\setlength{\tabcolsep}{4pt}
\begin{tabular}{cccccccccccccc}
\toprule
\multirow{3}{*}{KD Method} & \multirow{2}{*}{Model $\rightarrow$} & \multicolumn{3}{c}{VGG-16}    & \multicolumn{3}{c}{MobileNet-V2} & \multicolumn{3}{c}{ShuffleNet-V2} & \multicolumn{3}{c}{DenseNet-BC-121}  \\
                           &                                      & \multicolumn{3}{c}{(Teacher)} & \multicolumn{3}{c}{(Student A)}  & \multicolumn{3}{c}{(Student B)}   & \multicolumn{3}{c}{(Student C)}      \\
\cmidrule{3-14}
                           & Attack $\downarrow$                  & ASR   & BA    & TITG          & ASR   & BA    & TITG             & ASR   & BA    & TITG              & ASR   & BA    & TITG                 \\
\midrule
\multirow{2}{*}{Response}  & BackWeak (TO)                        & 4.43  & 89.36 & 2.77          & 3.44  & 90.04 & 1.40             & 2.84  & 89.71 & 0.71              & 3.12  & 91.34 & 1.37                 \\
                           & BackWeak                             & 99.89 & 86.14 & 87.45         & 99.30 & 88.90 & 88.97            & 98.90 & 88.52 & 88.17             & 99.42 & 90.21 & 88.63                \\
\midrule
\multirow{2}{*}{Feature}   & BackWeak (TO)                        & 4.43  & 89.36 & 2.77          & 3.34  & 84.95 & 1.09             & 4.04  & 83.37 & 0.53              & 3.83  & 88.36 & 1.49                 \\
                           & BackWeak                             & 99.89 & 86.14 & 87.45         & 90.20 & 81.91 & 78.16            & 53.50 & 81.29 & 45.48             & 99.29 & 86.15 & 86.71                \\
\midrule
\multirow{2}{*}{Relation}  & BackWeak (TO)                        & 4.43  & 89.36 & 2.77          & 3.46  & 89.21 & 1.45             & 3.33  & 88.71 & 0.99              & 3.36  & 89.11 & 1.20                 \\
                           & BackWeak                             & 99.89 & 86.14 & 87.45         & 99.86 & 87.98 & 88.12            & 99.58 & 87.03 & 87.18             & 99.84 & 86.62 & 87.40                \\
\bottomrule
\end{tabular}
\end{table*}

\begin{table*}[!t]
\centering
\caption{Performance (\%) of BackWeak on CIFAR-10 using the RegNetY architecture. \textbf{TO} denotes \emph{Trigger Only}, where triggered inputs are tested on \textbf{benign models} without any backdoor injection.}
\label{tab:regnet_results}
\setlength{\tabcolsep}{4pt}
\begin{tabular}{cccccccccccccc}
\toprule
\multirow{3}{*}{KD Method} & \multirow{2}{*}{Model $\rightarrow$} & \multicolumn{3}{c}{RegNetY-1.6GF} & \multicolumn{3}{c}{MobileNet-V2} & \multicolumn{3}{c}{ShuffleNet-V2} & \multicolumn{3}{c}{DenseNet-BC-121}  \\
                           &                                      & \multicolumn{3}{c}{(Teacher)}     & \multicolumn{3}{c}{(Student A)}  & \multicolumn{3}{c}{(Student B)}   & \multicolumn{3}{c}{(Student C)}      \\
\cmidrule{3-14}
                           & Attack $\downarrow$                  & ASR   & BA    & TITG              & ASR   & BA    & TITG             & ASR   & BA    & TITG              & ASR   & BA    & TITG                 \\
\midrule
\multirow{2}{*}{Response}  & BackWeak (TO)                        & 10.31 & 89.09 & 8.12              & 11.92 & 89.58 & 9.79             & 10.54 & 89.59 & 8.42              & 13.12 & 90.63 & 11.18                \\
                           & BackWeak                             & 99.88 & 87.31 & 90.96             & 96.63 & 89.28 & 87.82            & 94.43 & 88.63 & 85.90             & 98.14 & 90.25 & 89.07                \\
\midrule
\multirow{2}{*}{Feature}   & BackWeak (TO)                        & 10.31 & 89.09 & 8.12              & 11.02 & 84.58 & 8.33             & 6.19  & 82.72 & 2.78              & 11.89 & 88.42 & 9.75                 \\
                           & BackWeak                             & 99.88 & 87.31 & 90.96             & 96.78 & 82.04 & 87.37            & 65.40 & 79.96 & 59.14             & 99.88 & 86.43 & 89.93                \\
\midrule
\multirow{2}{*}{Relation}  & BackWeak (TO)                        & 10.31 & 89.09 & 8.12              & 12.82 & 89.24 & 10.61            & 5.81  & 88.36 & 3.29              & 9.64  & 89.25 & 7.97                 \\
                           & BackWeak                             & 99.88 & 87.31 & 90.96             & 97.40 & 88.63 & 89.04            & 57.18 & 87.20 & 52.61             & 96.03 & 88.02 & 87.95                \\
\bottomrule
\end{tabular}
\end{table*}

\end{document}